\documentclass[10pt, oneside, onecolumn]{article}

\usepackage[a4paper, margin=0.75in]{geometry}
\usepackage{cite}
\usepackage{float}
\usepackage{setspace}
\usepackage{multirow}
\usepackage{array}
\usepackage{import}
\usepackage{tikz} 
\usepackage[bookmarks=false]{hyperref}
\usepackage[hyphenbreaks]{breakurl} 
\usepackage{graphicx}
\usepackage{epstopdf}
\usepackage{xcolor,colortbl}
\usepackage{tabularx}
\usepackage{flushend}
\usepackage{amsmath}
\usepackage{subcaption}
\usepackage{keyval,xparse}

\begin{document}

\title{Averaging Gone Wrong: Using Time-Aware Analyses to Better Understand Behavior\footnote{A final version of this work was published in the Proceedings of the 25th International Conference on World Wide Web (WWW'16). It can be found in \url{http://dx.doi.org/10.1145/2872427.2883083}.}}

\author{
Samuel Barbosa\\
University of S\~ao Paulo\\
S\~ao Paulo, Brazil\\
sam@ime.usp.br
\and
Dan Cosley\\
Cornell University\\
Ithaca, NY 14853 USA\\
danco@cs.cornell.edu
\and
Amit Sharma\\
Microsoft Research\\
New York, NY 10011 USA\\
amshar@microsoft.com
\and
Roberto M. Cesar-Jr\\
University of S\~ao Paulo\\
S\~ao Paulo, Brazil\\
cesar@ime.usp.br
}

\date{10 October 2015}

\maketitle
\begin{abstract}
\looseness=-1
Online communities provide a fertile ground for analyzing people's behavior and improving our understanding of social processes.  Because both people and communities change over time, we argue that analyses of these communities that take time into account will lead to deeper and more accurate results.  Using Reddit as an example, we study the evolution of users based on comment and submission data from 2007 to 2014. Even using one of the simplest temporal differences between users---yearly cohorts---we find wide differences in people's behavior, including comment activity, effort, and survival.  Further, not accounting for time can lead us to misinterpret important phenomena.  For instance, we observe that average comment length decreases over any fixed period of time, but comment length in each cohort of users steadily increases during the same period after an abrupt initial drop, an example of Simpson's Paradox.  Dividing cohorts into sub-cohorts based on the survival time in the community provides further insights; in particular, longer-lived users start at a higher activity level and make more and shorter comments than those who leave earlier.  These findings both give more insight into user evolution in Reddit in particular, and raise a number of interesting questions around studying online behavior going forward.
\end{abstract}

\makeatletter
\define@key{subimagedic}{width}{\def\width{#1}}
\define@key{subimagedic}{scale}{\def\scale{#1}}
\makeatother
\NewDocumentCommand\subimage{O{} m G{}}{
    \begingroup
        \setkeys{subimagedic}{width={1.0},scale={1.0}, #1}
        \begin{subfigure}{\width\textwidth}
            \expandafter\includegraphics\expandafter[scale=\scale]{#2}
        \caption{#3}\end{subfigure}
    \endgroup
}

\section{Introduction}

Understanding the evolution of users in a social network is essential for a variety of tasks: monitoring community health, predicting individual user trajectories, and supporting effective recommendations, among others.  Many works aim at explaining these temporal aspects of evolution. Some adopt a point of view of the whole network and try to understand general patterns of behavior \cite{Zhu2014, Kooti2010}, while others adopt a user-centric point of view and try to model \cite{Correa2010, Priedhorsky2007, Panciera2009, Welser2011} or predict \cite{Danescu-niculescu-mizil2013} individuals' behavior.

These approaches often combine all available data into aggregate analyses of the whole community over its entire history.  This can be a natural response to limitations in the amount of available data:  datasets may capture a small part of the community's history \cite{Artzi2012}; timestamps may not be available \cite{Priedhorsky2007, Pujol2010}; snapshots may provide limited views of the community \cite{Cosley2010}; or the community itself may be small \cite{Lewis2008}.  Aggregate time-based analyses are also a natural first way to address questions of community evolution.

\looseness=-1
However, we argue that many of these aggregated views are misleading. The conditions under which users join the community may vary greatly over time in ways that might impact their behavior \cite{Miller2015}.  Among other things, popularity, purpose, features, interface, and algorithms can change: Wikipedia circa 2005 and circa 2015 are very different, as are Facebook of 2005 and 2015.  Analyses---including some of our own past work---that fail to account for this change may miss important details of what is really going on.

We support this argument through an analysis of user effort in Reddit, one of the most popular and long-running online communities, based on a very large, recently released dataset of posting behavior.  We address a number of questions commonly raised about users' effort in online communities: how active are users, how hard do they work, and what kinds of things do they do?  In each case, we compare aggregate analyses of posting behavior to ones that treat users in Reddit as yearly cohorts, and views that focus on calendar time versus user-referential views that normalize behavior based on the date of a user's first visible activity.  We also look at differences within yearly cohorts, focusing on how behavior differs between shorter and longer-lived users within each cohort.

We find that these accountings for time reveal insights about Reddit beyond what commonly performed aggregate analyses can provide.  Users who join Reddit earlier post more and longer comments than those who join later, while users who survive longer start out both more active and more likely to comment than submit versus users who leave Reddit early; none of these findings are obvious from aggregate views of user behavior.  

Further, we find that aggregate analysis can be downright misleading.  For instance, although average comment length decreases over time in an aggregate view, the comment length for surviving users increases over time in every cohort.  Likewise, an aggregate analysis suggests that longer-lived users post more over time; this is not the case.  Instead, users come into Reddit as active as they will ever be (akin to Panciera et al.'s finding that Wikipedians are ``born, not made'' \cite{Panciera2009}), and the rise in average activity for surviving users over time is driven by lower-activity users leaving early.

We see this paper as both making specific contributions to understanding behavior in Reddit and a more general contribution around the importance of considering change over time in analyzing online communities.

\section{Time matters} 

\subsection{Why accounting for time is important}

Communities grow and, with time, die. For any community, its users play a role in its evolution, but they are also simultaneously affected by the evolution of the community. Untangling this interplay can help make sense of patterns of activity in a community.

One useful way to understand the evolution of a community and its users is through time, as it provides a linear account of the growth (or decay) of overall activity, types of content, and social norms and structure.  One aspect of time often considered is the tenure of a user in the community, as in studies around modeling users' preferences \cite{McAuley2013} or analyzing the evolution of their language \cite{Danescu-niculescu-mizil2013}.  These analyses uncover insights about the lifecycle of a user in a community: users' preferences and behavior change with their age in a community \cite{Panciera2010}, while their early experiences and activity shape future outcomes predictably \cite{Tan2015,Yang2009,Panciera2009, Miller2015}. 

However, much past work on online communities ignores the time at which a user joins the community and analyzes all users together.
This might be a mistake: communities may grow denser or sparser with time \cite{Leskovec2005}, develop new norms \cite{Kooti2010}, and enact policies and rules guiding people's behavior \cite{Butler2008}.
These changes mean that people experience different versions of a community at different times, which can, in turn, affect their observed behavior. This interaction with the state of a community can confound conclusions about people's behavior, because the differences one observes may simply due to changes in the community, rather than any significant change in the outcome variable of interest or the user population.

\subsection{Cohorts are analytically useful}

A common method to control for such confounds is cohort analysis, widely used in fields such as sociology \cite{Mason2012,Glenn2005}, economics \cite{Attanasio1993,Beldona2005}, and medicine \cite{Howartz1996,Davis2010}. A cohort is defined as a group of people who share a common characteristic, generally with respect to time. For example, people born in the same year, or those who joined a school at the same time, or got exposed to an intervention at similar times can be considered as cohorts.  People in a cohort are assumed to be exposed to the same state of the world and thus are more comparable to each other than to people in other cohorts. 

For example, sociological studies often use students who join a school in the same year to understand the effect of interventions \cite{Goyette2008,Alexander2012}, and condition on the year in which people were born to understand people's  behavior, such as variations in financial decision-making \cite{Attanasio1993} or opinions on issues \cite{Firebaugh1988,Jennings1996}. Similarly, medical studies interpret effects of drugs using cohorts of people within the same age group or amount of exposure to correlated conditions \cite{Howartz1996,Davis2010}.  

Recent work shows that cohorts' importance transfers to online communities as well. Just as people's behavior varies according to their biological age, their experience in an online community may vary with their age in the community and their year of joining. In Wikipedia, we find substantial differences in the activities of cohorts of users who joined earlier versus those who joined later \cite{Welser2011}. Similarly, on review websites, users who join later tend to adopt different phrases than the older users who had joined earlier \cite{Danescu-niculescu-mizil2013}.

\subsection{What might cause these differences?}

These differences in activity between cohorts may be due to a number of reasons.  One plausible explanation is selection effects: people who are enthusiastic about a community or its goals are more likely to self-select as early members of a community, while others may be more likely to join later \cite{Li2008}.  In this case, users who join earlier might be expected to be more active, committed users than those who join later. 

Another possible explanation is that community norms may change over time.  In many cases, it is a bottom-up process. Kooti et al. showed that social conventions can define the evolution of a community and the early adopters play a major role in designing these conventions, consciously or not \cite{Kooti2010}. Examples include adoption of `RT', a retweeting norm by Twitter users and the subsequent introduction of the Retweet button on Twitter \cite{Kooti2010}; change in language use between new and old users on review websites \cite{Danescu-niculescu-mizil2013}; and assumptions of clear roles and responsibilities on Wikipedia \cite{Kittur2007a}. In other cases, it may be directed by the community managers. For instance, the makers of Digg unilaterally changed the nature of the community by introducing a new version of the website, leading to a sudden change in norms and behavior in the community \cite{Ingram2014,Lardinois2014}. 

The growth of a community may also affect people's behavior. Successful communities often grow very rapidly, which can be both good and bad for users' experience. On one hand, growth would imply availability of a larger chunk of content to choose from. On the other, it might be harder to connect to others and get responses in a bigger community. A community may also need to adopt new rules and policies to manage growth and newcomers, as in the evolution of Wikipedia \cite{Choi2010,Bryant2005}. In those cases, the experience of later cohorts of users may be vastly different from the initial ones who joined before formal rules were in place. 

Finally, patterns of use may change because the overall population of Internet users is still changing.  As more and different people come online, their influx may lead to changes in activity patterns and communities (as with the yearly entry of college freshmen, and eventually all of AOL, gaining access to Usenet).  The gradual penetration of technology also has age-related effects:  people who did not grow up in a technological environment differ in their social media and search usage compared to younger generations\cite{Correa2010,Beldona2005}. 

\subsection{Is Reddit getting ``worse'' over time?}
All of the above reasons suggest that users from different cohorts are likely to be different, which has also been demonstrated in online and offline communities \cite{Ryder1965,Danescu-niculescu-mizil2013,Prensky2001,Correa2010}.  Further, they suggest a general hypothesis that communities ``get worse'' over time because newer users are likely to be less committed and knowledgeable about the community. 

To address this hypothesis, we analyze both aggregate and cohort-based measures of user quality that are often raised about online communities: how active are users \cite{Scellato2011,Hughes2009,Java2007,Levy1984}, how much do they contribute \cite{Scellato2011,Gruhl2004,Guo2009}, and what kinds of work do they engage in \cite{Welser2011,Choi2010,Panciera2009}?  

We do this in the context of Reddit, a community that has been studied by many researchers \cite{Gilbert2013,Stoddard2015,Bergstrom2011,Tan2015}. We begin with a brief overview of both Reddit and the dataset that we use in this paper, focusing on aspects that directly impact our analyses\footnote{There is more to say about Reddit itself (see \cite{AboutReddit}).}.

\section{Data: Reddit as a community}

\subsection{What is Reddit, briefly}

\begin{figure}[!tb]
\centering
\includegraphics[width=0.45\textwidth,natwidth=964,natheight=823]{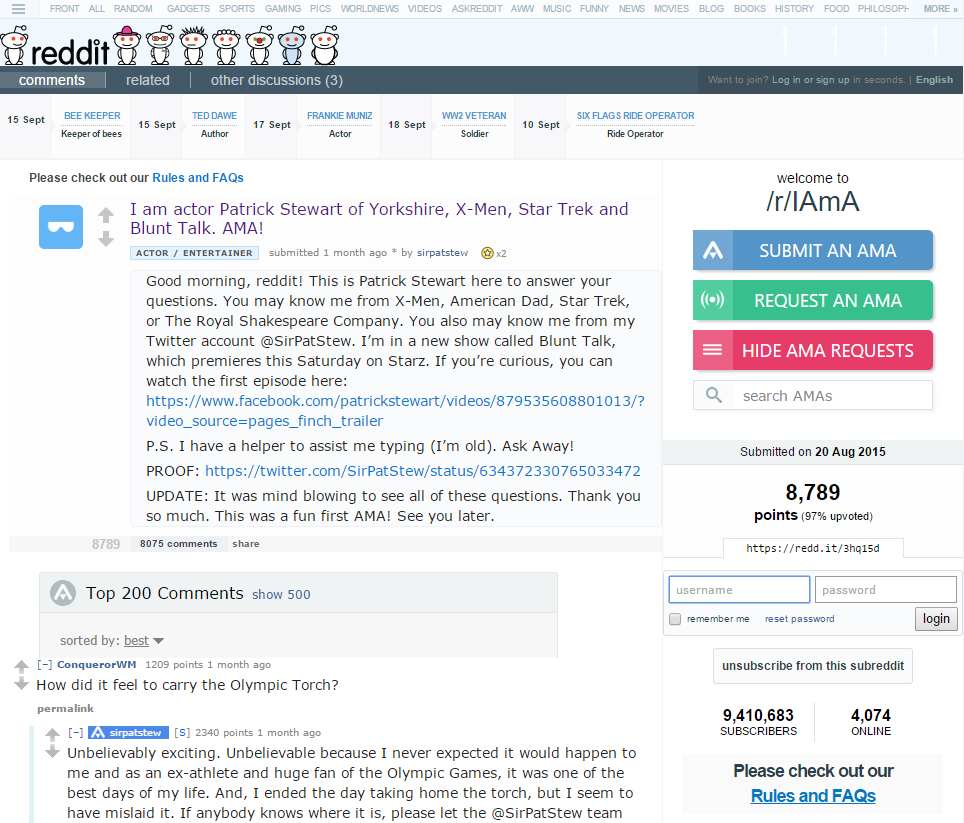}
\caption{Reddit interface when visualizing a submission. This is Patrick Stewart's ``AmA'' (ask me anything) in ``IAmA'' (I am a), a submission where he answers users' questions in the comments. We can see the most upvoted comment and Patrick's answer right below.}
\label{fig:reddit}
\end{figure}

Reddit is one of the largest sharing and discussion communities on the Web.  According to Alexa, as of late 2015 Reddit is in the top 15 sites in the U.S. and the top 35 in the world in terms of monthly unique visitors.  It consists of a large number of subreddits (853,000 as of June 21st, 2015\footnote{\cite{RedditStatistics} provides more statistics about Reddit.}), each of which focuses on a particular purpose.  Many subreddits are primarily about sharing web content from other sites: in ``Pics'', ``News'', ``Funny'', ``Gaming'', and many other communities, users (``Redditors'') make ``submissions'' of links posted at other sites that they think are interesting.  In other subreddits, Redditors primarily write text-based ``self-posts'': ``AskReddit'', ``IAmA'', and ``ShowerThoughts'' are places where people can ask questions and share stories of their own lives.  Generically, we will refer to submissions and text posts as ``submissions''.  

Each submission can be imagined as the root of a threaded comment tree, in which Redditors can comment on submissions or each other's comments.  Redditors can also vote on both submissions and comments; these votes affect the order in which submissions and comments are displayed and also form the basis of ``karma'', a reputation system that tracks how often people upvote a given Redditor's comments and submissions. We can observe these elements in Figure~\ref{fig:reddit}. 

We choose Reddit as our target community for a number of reasons.  It has existed since 2005, meaning that there has been ample time for the community to evolve and for differences in user cohorts to appear.  Second, it is one of the most popular online communities, allowing different types of contributions---comments and original submissions---across many different subreddits.  Third, a number of Reddit users believe that it is, in fact, getting worse over time\cite{RedditWorse1,RedditWorse2,RedditWorse3,RedditWorse4,RedditWorse5,RedditWorse6}. Finally, Reddit data are publicly available through an API.

\subsection{The dataset}

\begin{figure*}[!tb]
\centering
\subimage[width=0.48, scale=0.40]{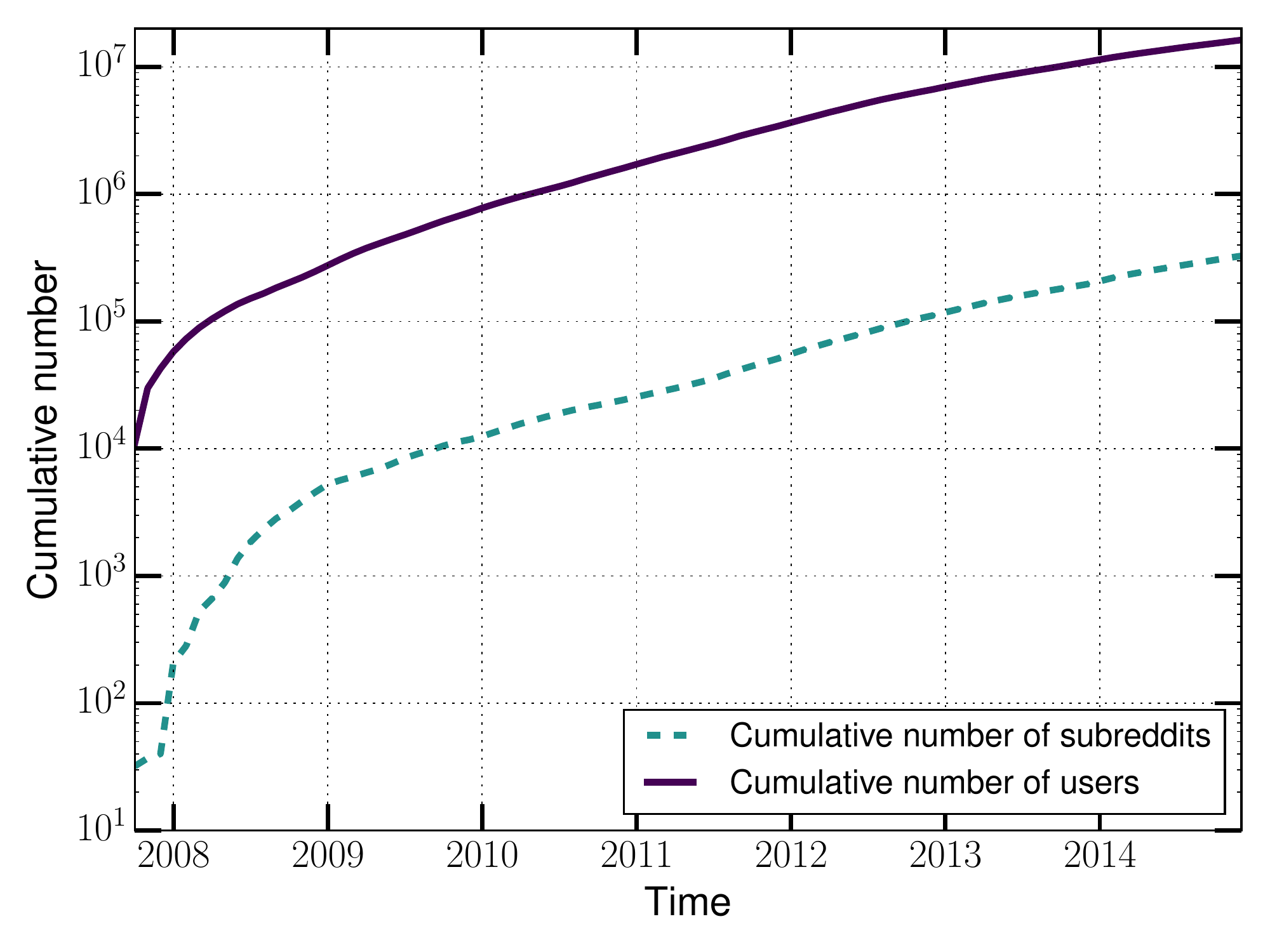}
\subimage[width=0.48, scale=0.40]{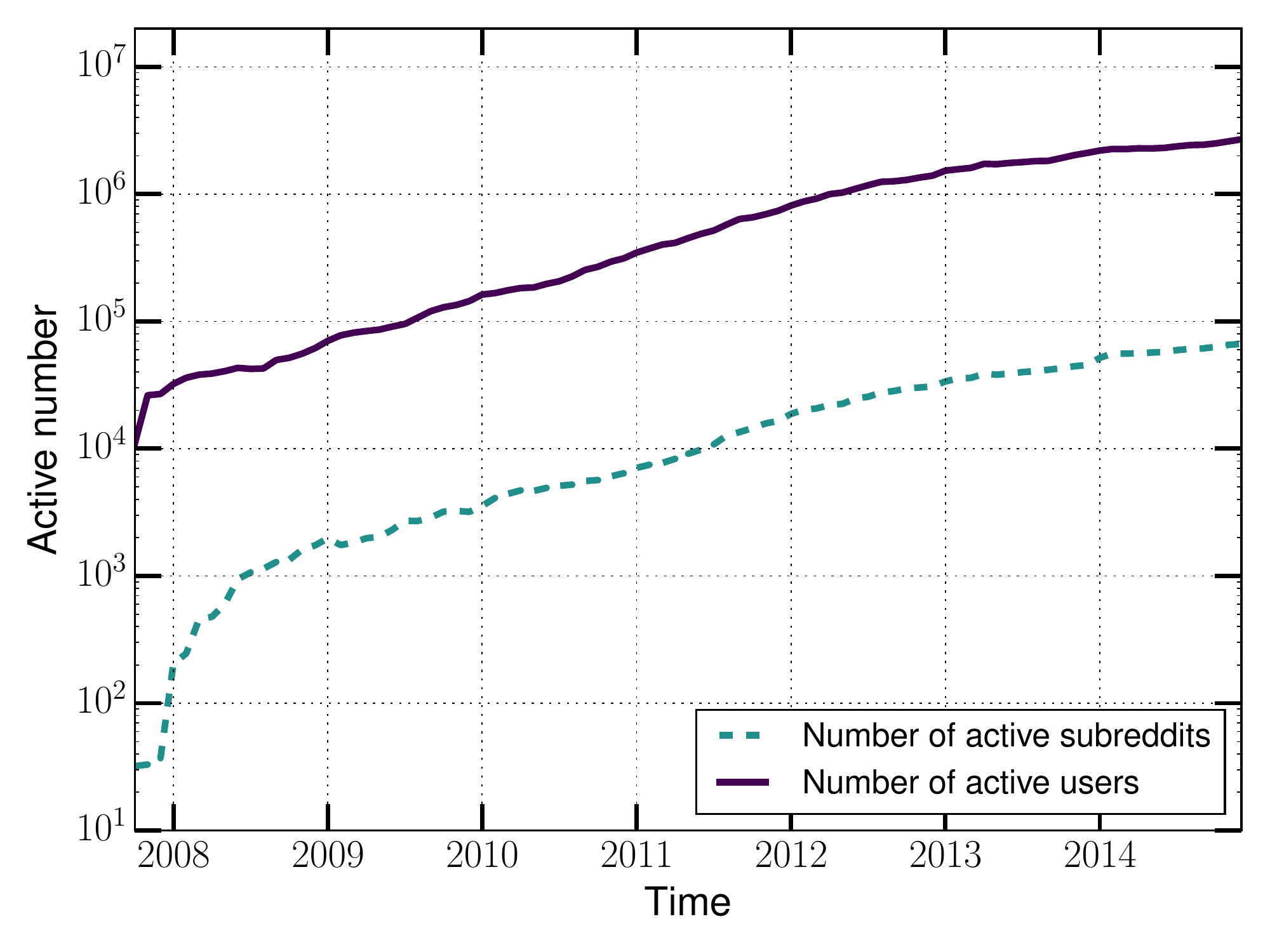}
\caption{Figure (a) shows the cumulative growth of Reddit for users and subreddits. Figure (b) shows the number of active users and subreddits in Reddit over time. An active user or subreddit is one that had at least one post (comment or submission) in the time bin we used---here, discretized by month.}
\label{fig:cumulative}
\end{figure*}

Redditor \textit{Stuck\_In\_The\_Matrix} used Reddit's API to compile a dataset of almost every publicly available comment\cite{RedditDataset1} from October 2007 until May 2015.  The dataset is composed of 1.65 billion comments, although due to API call failures, about 350,000 comments are unavailable.  He also compiled a submissions dataset for the period of October 2007 until December 2014 (made available for us upon request) containing a total of 114 million submissions.  These datasets contain the JSON data objects returned by Reddit's API for comments and submissions\footnote{A full description of the JSON objects is available at \cite{RedditAPI}.}; for our purposes, the main items of interest were the UTC creation date, the username, the subreddit, and for comments, the comment text.

\looseness=-1
We focus on submissions and comments in the dataset because they have timestamps and can be tied to specific users and subreddits, allowing us to perform time-based analyses.   In some analyses, we look only at comments; in some, we combine comments and submissions, calling them \textbf{``posts''}.  We would also like to have looked at voting behavior as a measure of user activity\footnote{This would also give us more insight than usual into lurkers' behavior; we'll return to this in the discussion.}, but individual votes with timestamps and usernames are not available through the API, only the aggregate number of votes that posts receive.

\subsection{Preprocessing the dataset}

\looseness=-1
To analyze the data, we used Google BigQuery\cite{BigQuery}, a big data processing tool.
Redditor \textit{fhoffa} imported the comments into BigQuery and made them publicly available\cite{RedditDataset2}.  We uploaded the submission data ourselves using Google's SDK.

For the analysis in the paper, we did light preprocessing to filter out posts by deleted users, posts with no creation time, and posts by authors with bot-like names\footnote{Ending with ``\_bot'' or ``Bot''; or containing ``transcriber'' or ``automoderator''.}.

We also considered only comment data from October 2007 until December 2014 in order to have a matching period for comments and submissions. After this process, we had a total of 1.17 billion comments and 114 million submissions.

\subsection{An overview of the dataset}

Here we present an overview of the dataset that shows Reddit's overall growth.  Figure~\ref{fig:cumulative}a presents the cumulative number of user accounts and subreddits created as of the last day of every month. After an initial extremely rapid expansion from 2008--2009, the number of users and subreddits have grown exponentially.  As of the end of 2014, about 16.2 million distinct users and 327 thousand subreddits made/received at least one post based on our data.

However, as with many other online sites, most users \cite{Scellato2011,Hughes2009,Java2007} and communities \cite{Arguello2006} do not stay active. We define as an ``\textbf{active user}'' one that made at least one post in the month in question. Similarly, an ``\textbf{active subreddit}'' is one that received at least one post in the month. In December 2014, about 2.7 million users and 66 thousand subreddits were active, both around a fifth of the cumulative numbers. Figure~\ref{fig:cumulative}b shows the monthly number of active users and subreddits.

Our interest in this paper is not so much whether users survive as it is about the behavior of active users.  Thus, 
in general our analysis will look only at active users and subreddits in each month; those that are temporarily or permanently gone from Reddit are not included.  

\subsection{Identifying cohorts}

\begin{figure*}[!tb]
\centering
\subimage[width=0.48, scale=0.40]{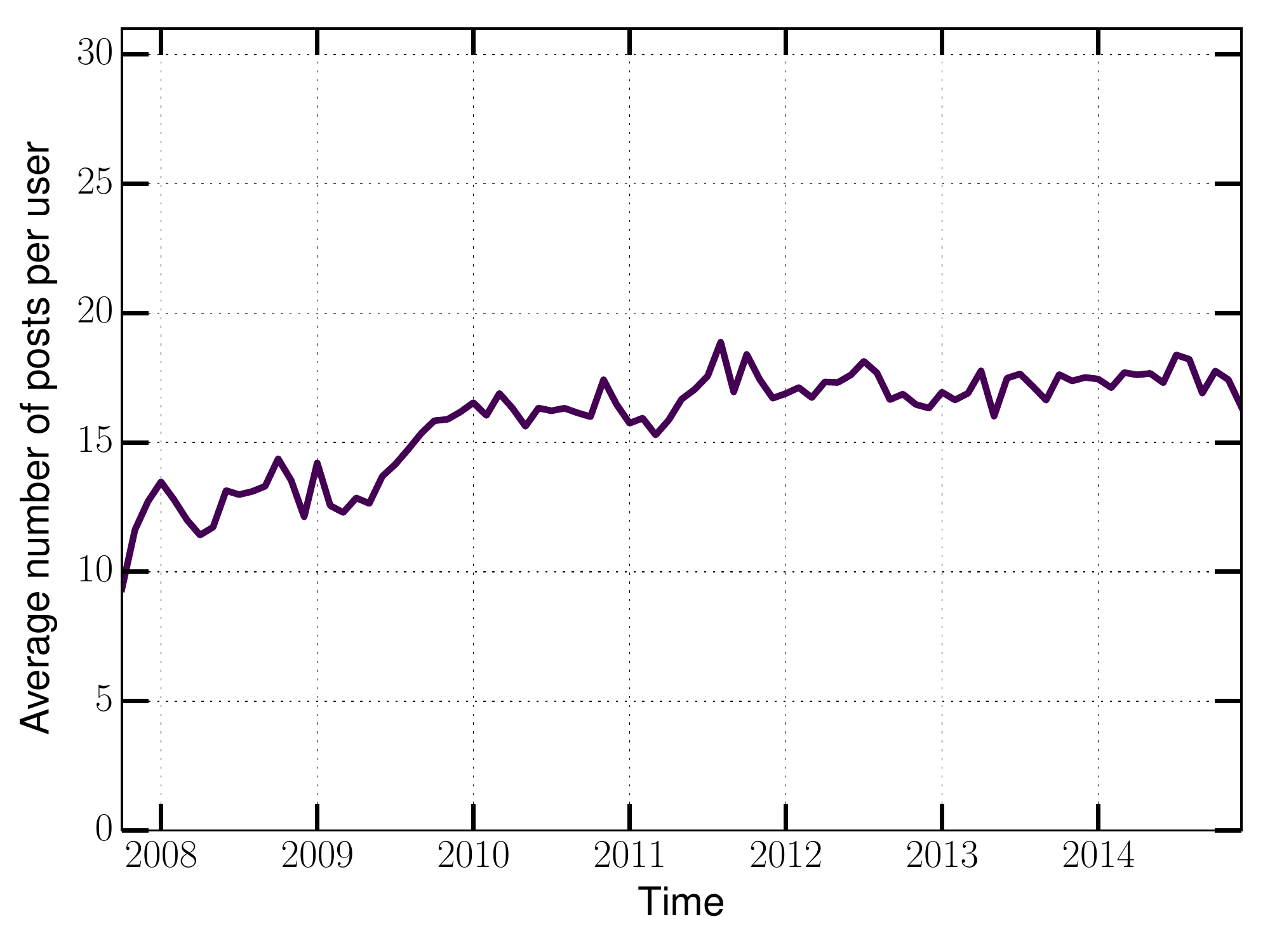}
\subimage[width=0.48, scale=0.40]{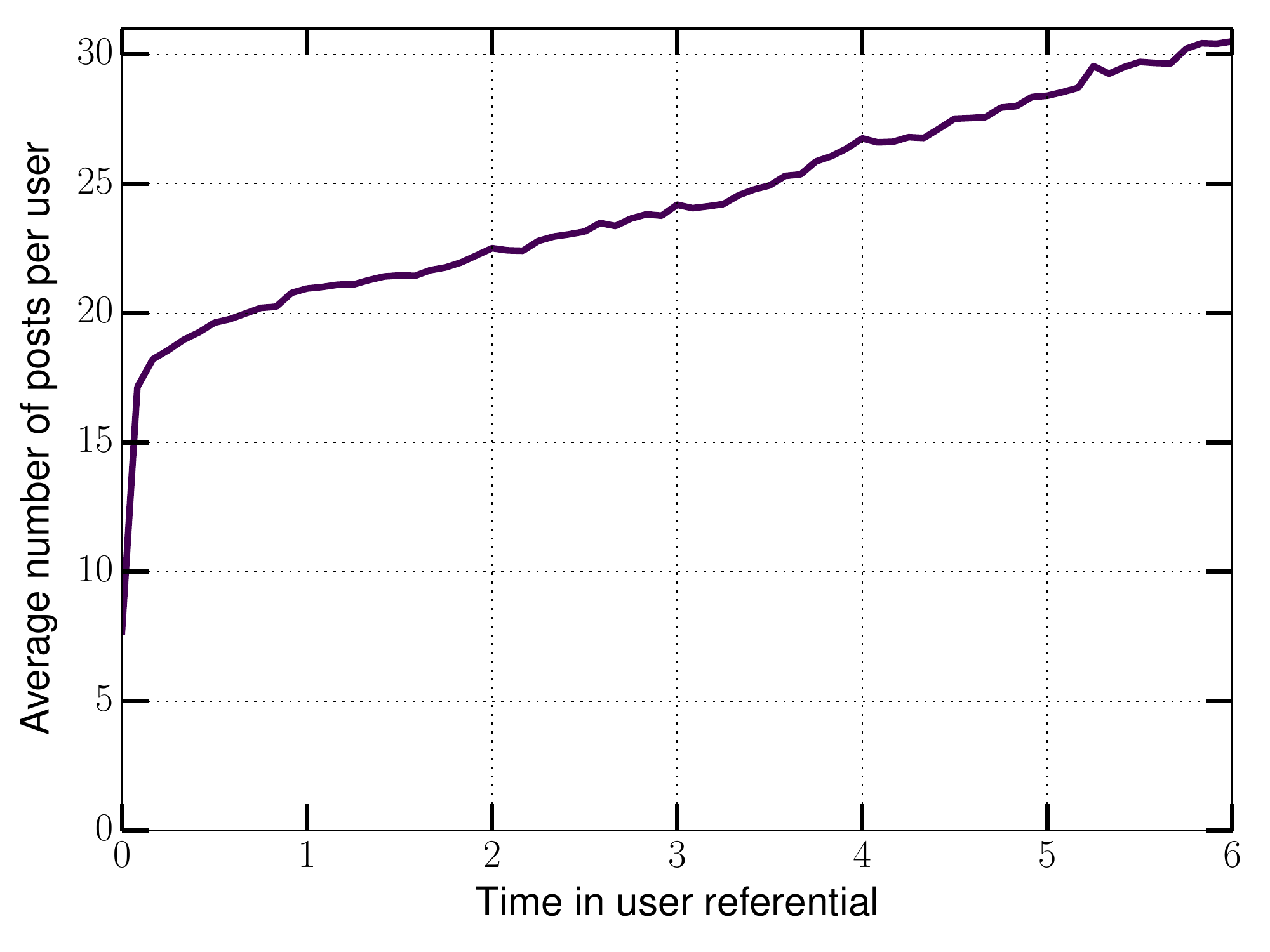}
\caption{In Figure (a), monthly average posts per active user over clock time. In Figure (b), monthly average posts per active users in the user-time referential, i.e., message creation time is measured relative to the user's first post.  Each tick in the x-axis is one year.  In both figures (and all later figures), we consider only active users during each month; users that are either temporarily or permanently away from Reddit are not included.}
\label{fig:overall_posts}
\end{figure*}

We define the ``\textbf{user's creation time}'' as the time of the first post made by that user.  Throughout this paper, we will use the notion of user cohorts, which will consist of users created in the same calendar year.

In many cases, we will look at the evolution of these cohorts. Since users can be created at any time during their cohort year, and our dataset ends in 2014, 
we are likely to have a variation on the data available for each user of up to one year, even though they are in the same cohort.  To deal with this, some of our cohorted analyses will consider only the overlapping time window for which we collect data for all users in a cohort.   This means that we are normally not going to include the 2014 cohort in our analyses.

Our data starts in October 2007, but Reddit existed before that. That means that, not only do we have incomplete data for the 2007 year (which compromises this cohort), but there might also be users and subreddits that show up in 2007 that were actually created in the previous years. Since we can not control for these, we will also omit 2007 cohort. We will, however, include 2007 in the overall analyses over time (the non cohorted ones) for two reasons: first, it does not have any direct impact on the results; second, we often compare the cohorted approach with a naive approach based on aggregation, and we would not expect a naive approach to do such filtering.

\section{Average posts per user}
One common way to represent user activity in online communities is quantity: the number of posts people make over time. Approaches that consider the total number of posts per user in a particular dataset \cite{Gruhl2004} and that analyze the variation of the number of posts per user over time \cite{Guo2009} have been applied to online social networks.  In this section, we use this measure to address our first research question (\textbf{RQ1)}: how does the amount of users' activity change over time?

As we will see, both visualizing behavior relative to a user's creation time and using cohorts provide additional insight into posting activity in Reddit compared to a straightforward aggregate analysis based on calendar time.

\subsection{Calendar versus user-relative time}

Figure~\ref{fig:overall_posts}a shows that aggregate analysis, presenting the average number of posts per month by active users in that month.  Taken at face value, this 
suggests that over the first few years of Reddit, users became more active in posting, with per-user activity remaining more or less steady since mid-2011.

\looseness=-1

This average view hides several important aspects of users' activity dynamics. Previous work has looked into behavior relative to the user creation time. It has been shown that edge creation time in a social network relative to the user creation follows an exponential distribution \cite{Tomkins2008}. User lifetime, however, does not follow a exponential distribution and some types of user content generation follow a stretched exponential distribution \cite{Guo2009}. Throw-away accounts are one example of very short-lived users in Reddit \cite{Bergstrom2011}, for example. 

To address these characteristics, Figure~\ref{fig:overall_posts}b shows a view that emphasizes the trajectory over a user's lifespan rather than the community's.  To do this, we scale the x-axis not by clock time, as in Figure~\ref{fig:overall_posts}a, but by time since the user's first post: ``1'' on the x-axis refers to one year since the user's account first post, and so on. We call this the \textbf{time in the user referential}. One caution about interpreting graphs with time in the user referential is that the amount of data available rapidly decreases over time as users leave the community, meaning that values toward the right side of an individual data series are more subject to individual variation.  

The evidence at this point supports the tempting hypothesis that the longer a user survives, the more posts they make (\textbf{H1}).  This hypothesis, however, is incorrect; we will present a more nuanced description of what is happening informed by cohort-based analyses.

\subsection{New cohorts do not catch up}

\begin{figure*}[!tb]
\centering
\subimage[width=0.48, scale=0.42]{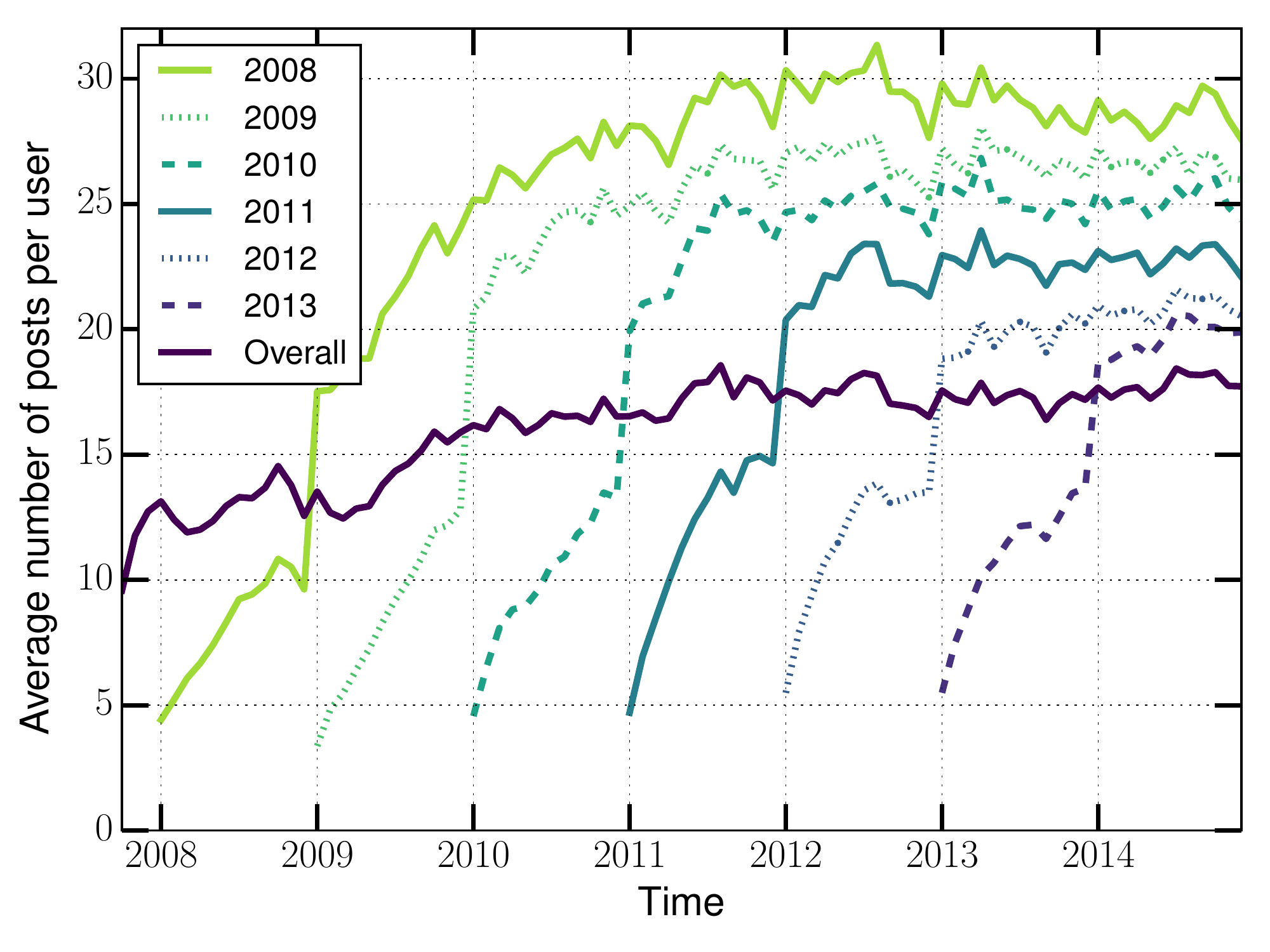}
\subimage[width=0.48, scale=0.42]{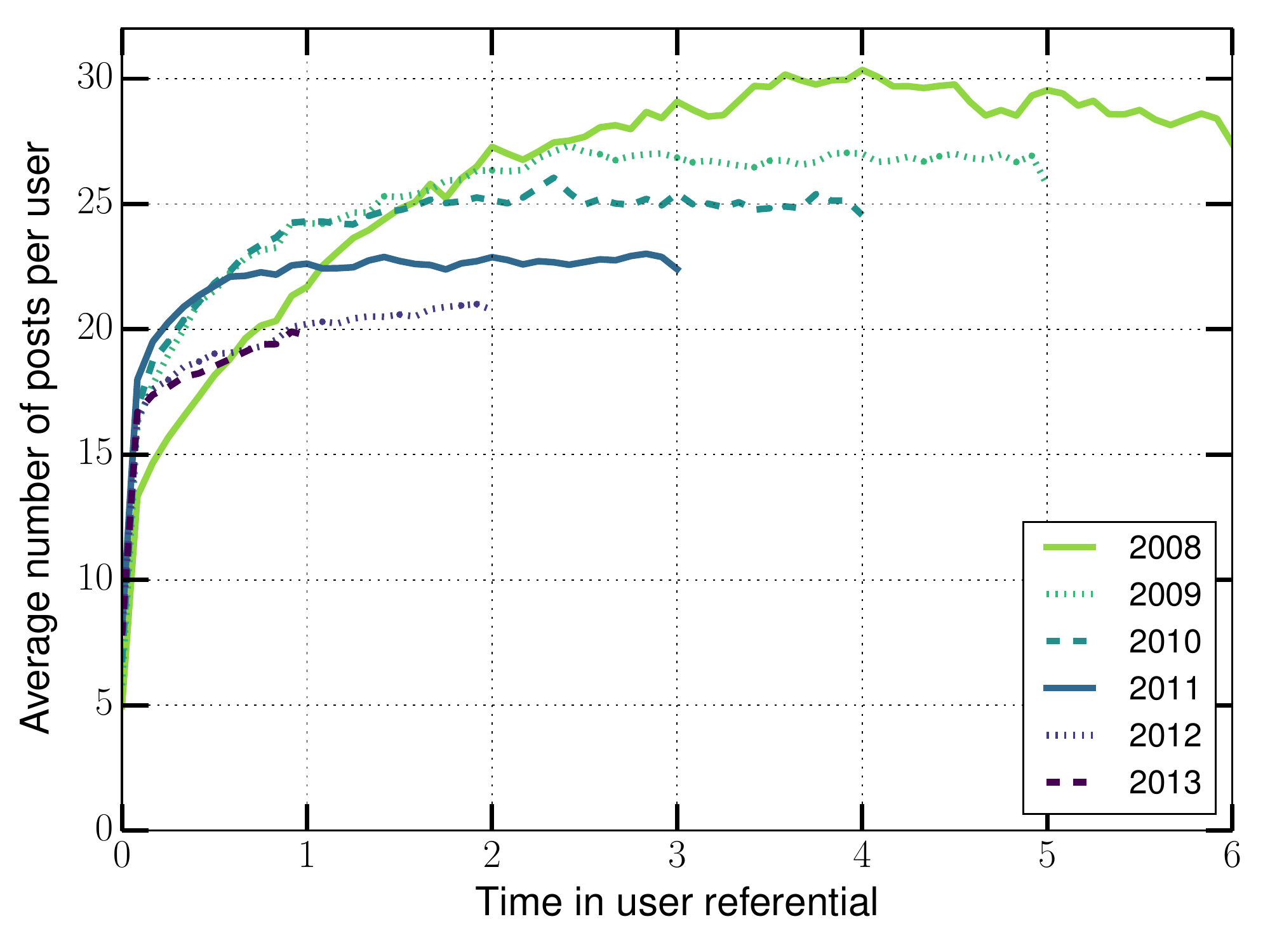}
\caption{Figure (a) shows the average number of posts per active user over clock time and Figure (b) per active user in the user-time referential, both segmented by users' cohorts. The user cohort is defined by the year of the user's creation time.  For comparison, the black line in Figure (a) represents the overall average.}
\label{fig:avr_posts_per_user_over_time_cohorts}
\end{figure*}

Figure~\ref{fig:overall_posts}b suggests that older users are more active than newer ones, raising the question of whether new users
eventually follow in older users' footsteps (\textbf{RQ1a}).  

Analyzing users' behavior by cohort is a reasonable way to address this question, and Figure~\ref{fig:avr_posts_per_user_over_time_cohorts}a shows a first attempt at this analysis.  We can already observe a significant cohort effect: users from later cohorts appear to level off at significantly lower posting averages than users from earlier ones.  It suggests that newer users likely will never be as active as older ones on average.  It also shows that surviving users are significantly more active than the overall average (the black line in the figure) would suggest.

However, Figure~\ref{fig:avr_posts_per_user_over_time_cohorts}a also has an awkward anomaly: a rapid rise in the average number of posts during each cohort's first calendar year, especially in December. Combining cohort segmentation with user-referential analysis, as in Figure~\ref{fig:avr_posts_per_user_over_time_cohorts}b, helps smooth out this anomaly and aligns cohorts with each other.  Doing this alignment makes clear that differences between earlier and later cohorts are apparent early on.

\subsection{Does tenure predict activity, or vice versa?}

\begin{figure*}[!tb]
\centering
\subimage[width=0.31, scale=0.29]{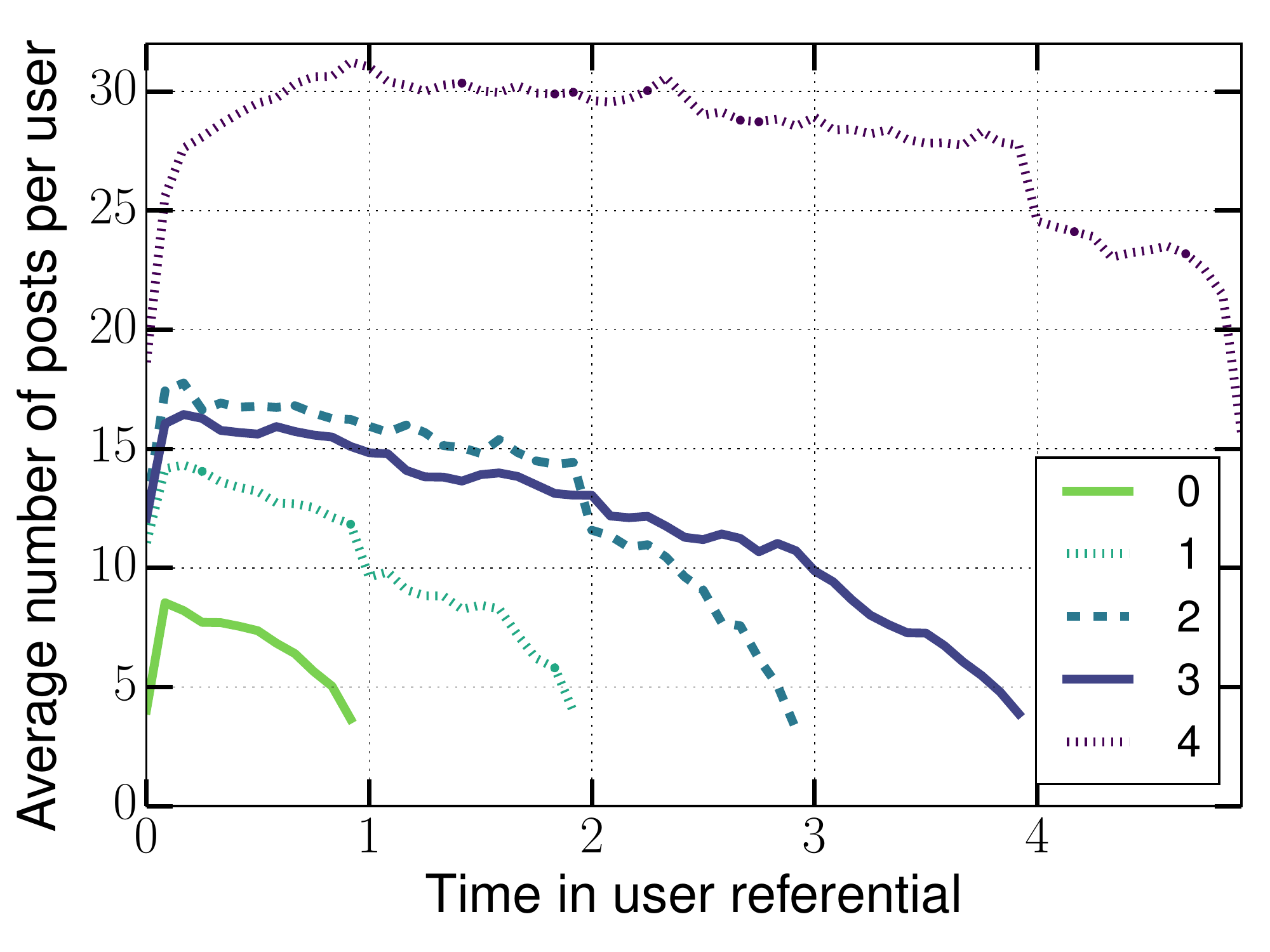}{2010 cohort}
\subimage[width=0.31, scale=0.29]{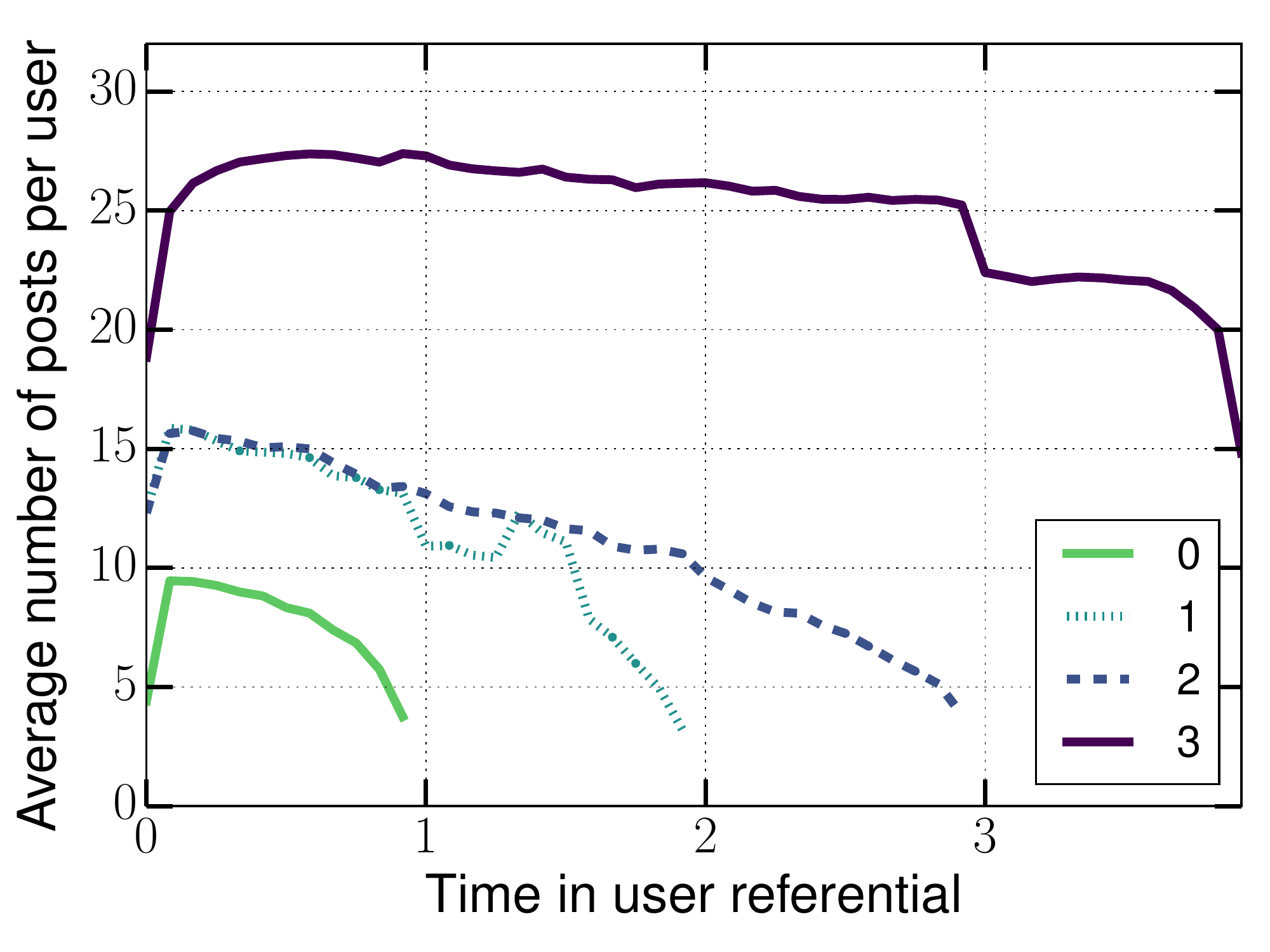}{2011 cohort}
\subimage[width=0.31, scale=0.29]{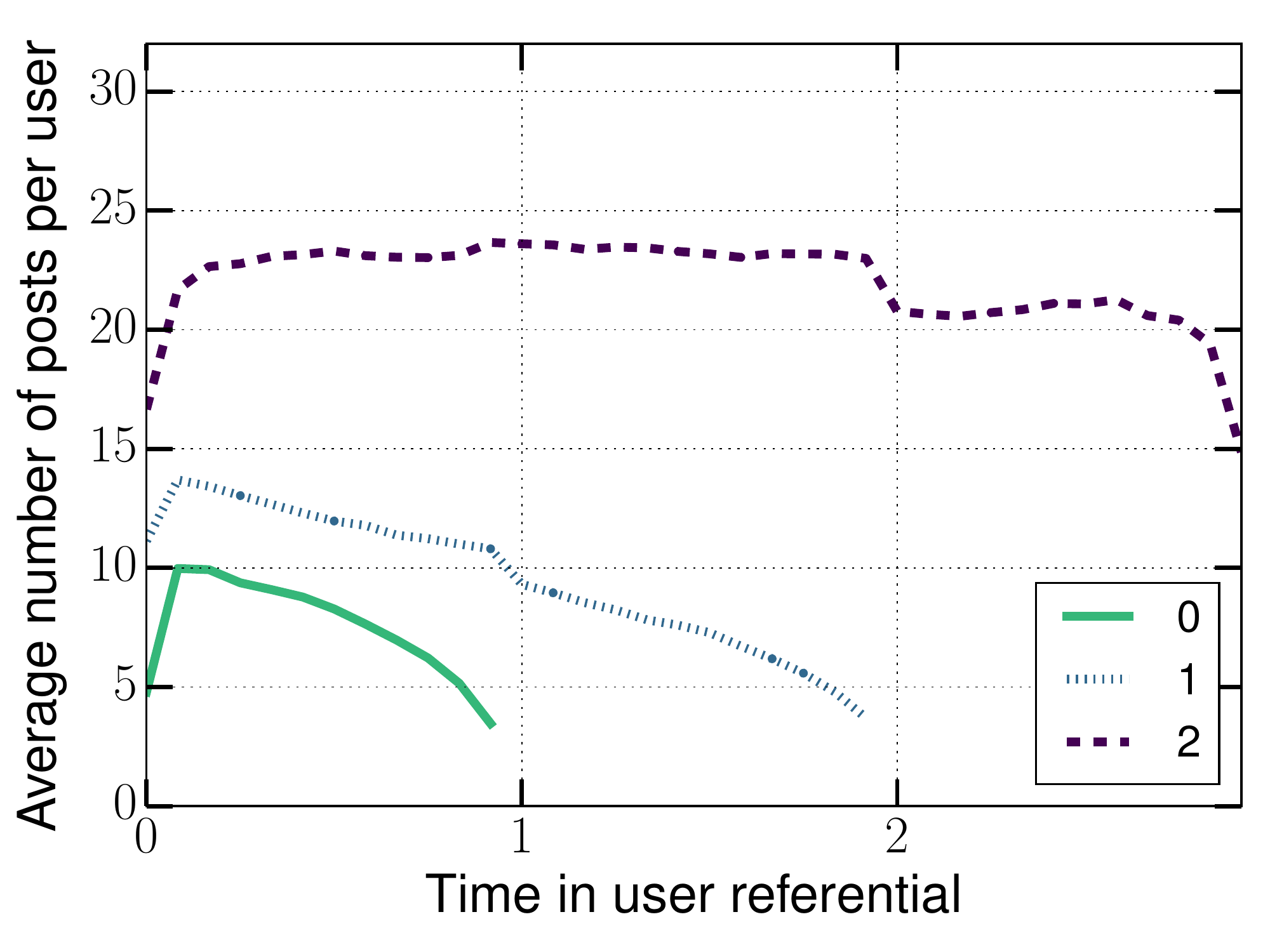}{2012 cohort}
\caption{Each Figure corresponds to one cohort, from 2010 to 2012, left to right. The users for each cohort are further divided in groups based on how long they survived: users that survived up to 1 year are labeled 0, from 1 to 2 years are labeled 1, and so on.  For all cohorts, longer-tenured users started at higher activity levels than shorter-tenured ones.}
\label{fig:avr_posts_per_user_for_surviving_year}
\end{figure*}

\looseness=-1
These graphs still support our initial hypothesis \textbf{H1} 
and they do not explain the rapid increase in posting activity in the first few months.  An alternative hypothesis, inspired by the ``Wikipedians are Born, not Made'' paper \cite{Panciera2009}, is that individual users come in with different posting propensities, and the rise over time is not that individual users become more active but that low-activity users leave the system (\textbf{H2}).  To examine this, we further segment each cohort by the number of years they were active in the system, as defined by the difference between their first and last post times.
 
Figure~\ref{fig:avr_posts_per_user_for_surviving_year} shows this analysis for the 2010, 2011 and 2012 cohorts\footnote{We only show these figures for the sake of saving space, but the same trends are observed in the other cohorts.}.  Across all cohorts and yearly survival sub-cohorts, users who leave earlier come in with a lower initial posting rate.  Thus, the rise in average posts per active user is driven by the fact that users who have high posting averages throughout their lifespan are the ones who are more likely to survive.  As the less active users leave the system, the average per active user increases.  In other words, the correct interpretation of Figure~\ref{fig:overall_posts}b is not \textbf{H1}: longer-lived users don't post more as they age.  Instead, users who post more---right from the beginning---live longer, supporting (\textbf{H2}). 

Combining Figure~\ref{fig:avr_posts_per_user_for_surviving_year}'s insight that the main reason why these curves increase is because the low posting users are dying sooner with the earlier observation that the stable activity level is lower for newer cohorts suggests that low-activity users from later cohorts tend to survive longer than those from earlier cohorts.  That is, people joining later in the community's life are less likely to be either committed users or leave than those from earlier on: they are more likely to be ``casual'' users that stick around.

\vspace{7pt} 
\section{Comment length}

\begin{figure*}[!tb]
\centering
\subimage[width=0.48, scale=0.42]{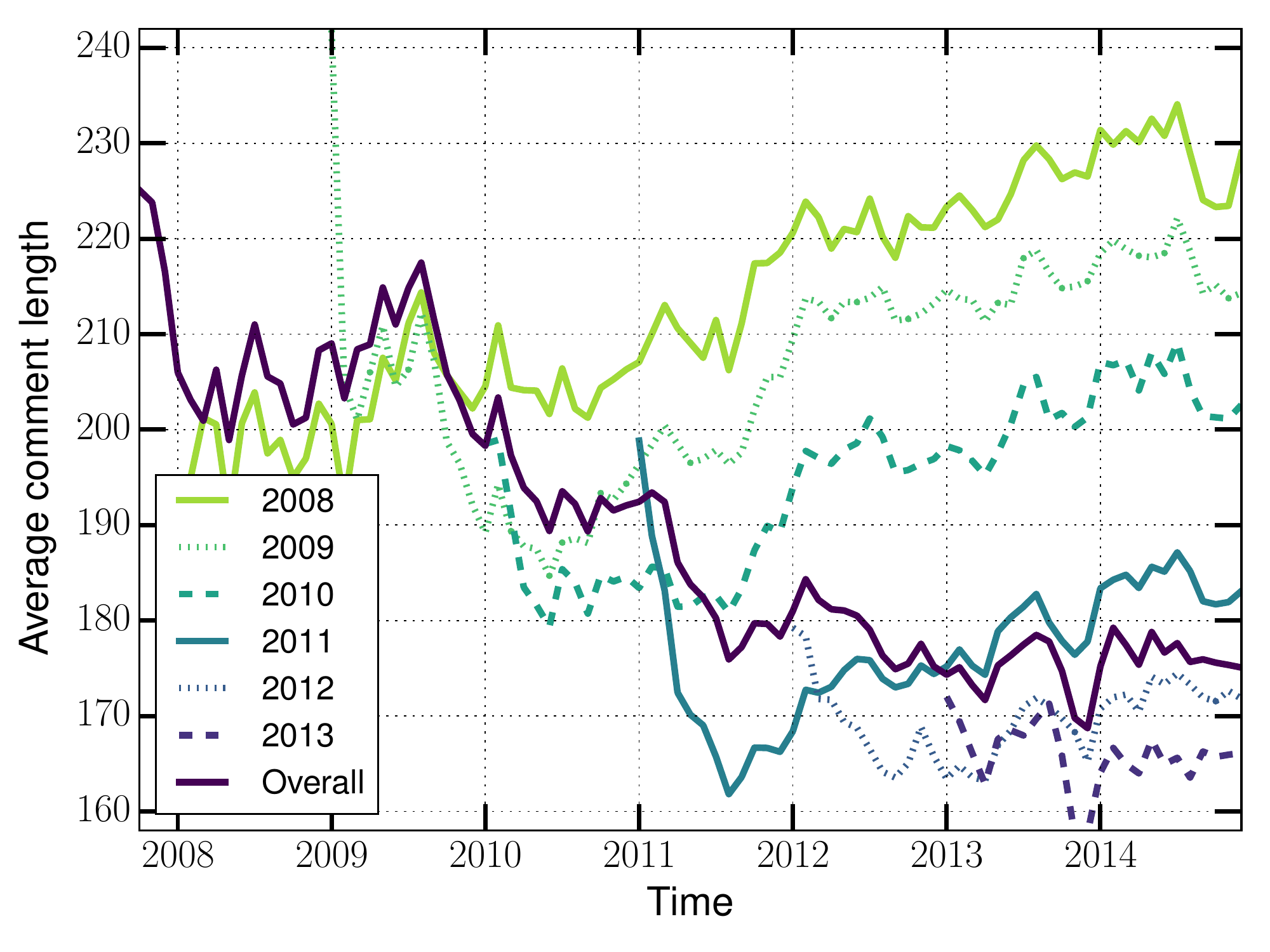}
\subimage[width=0.48, scale=0.42]{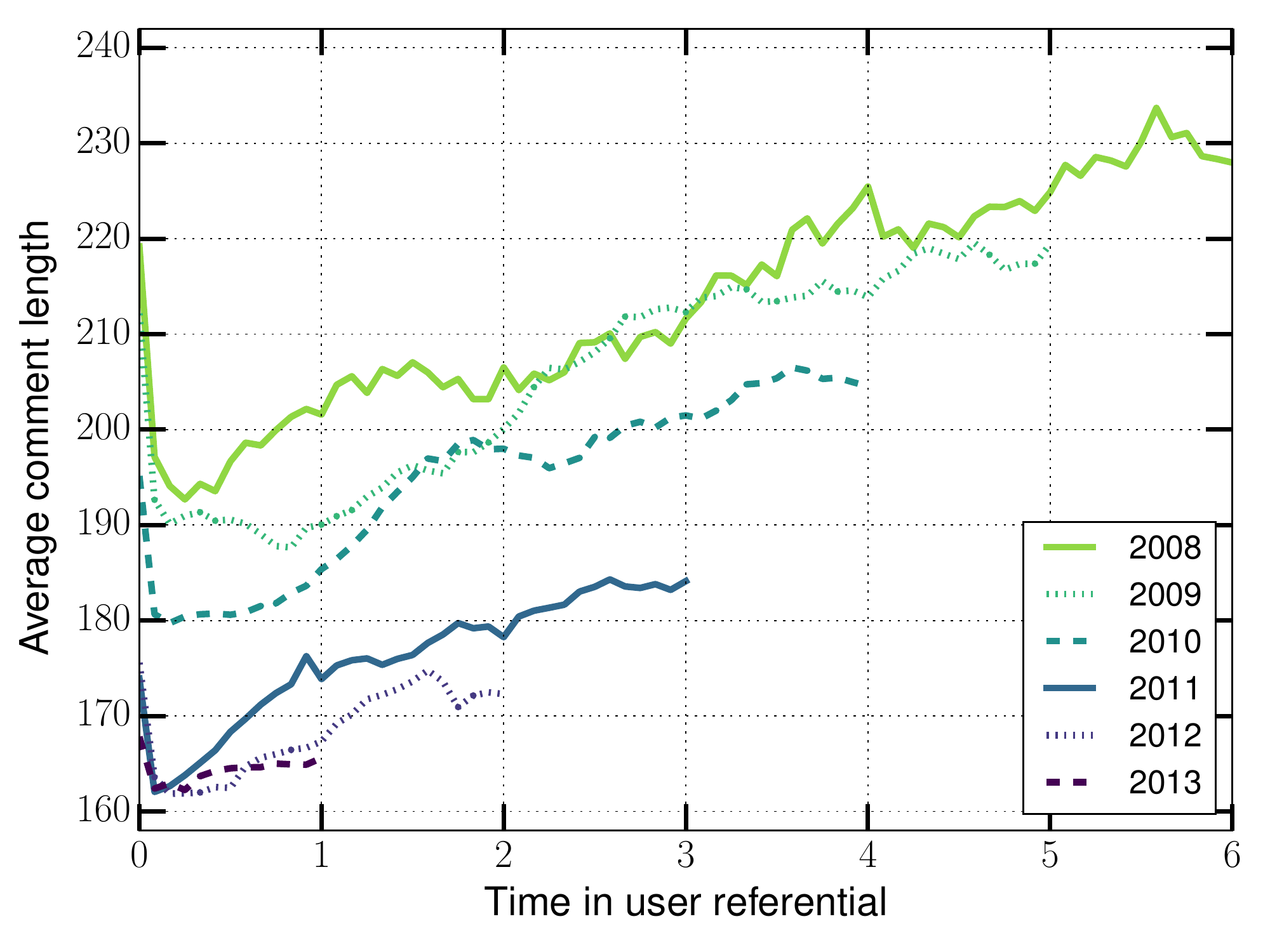}
\subimage[width=0.31, scale=0.29]{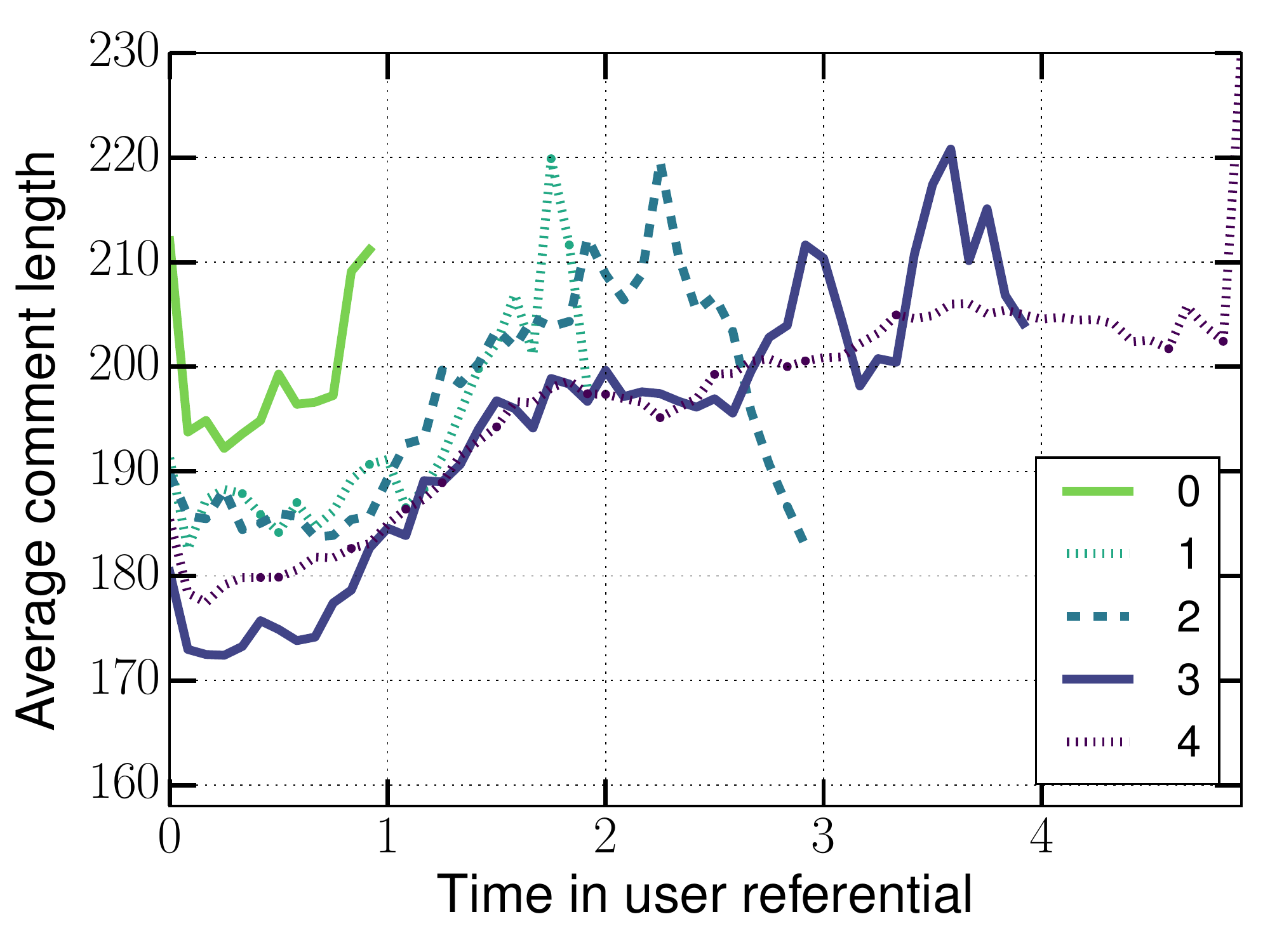}{2010 cohort}
\subimage[width=0.31, scale=0.29]{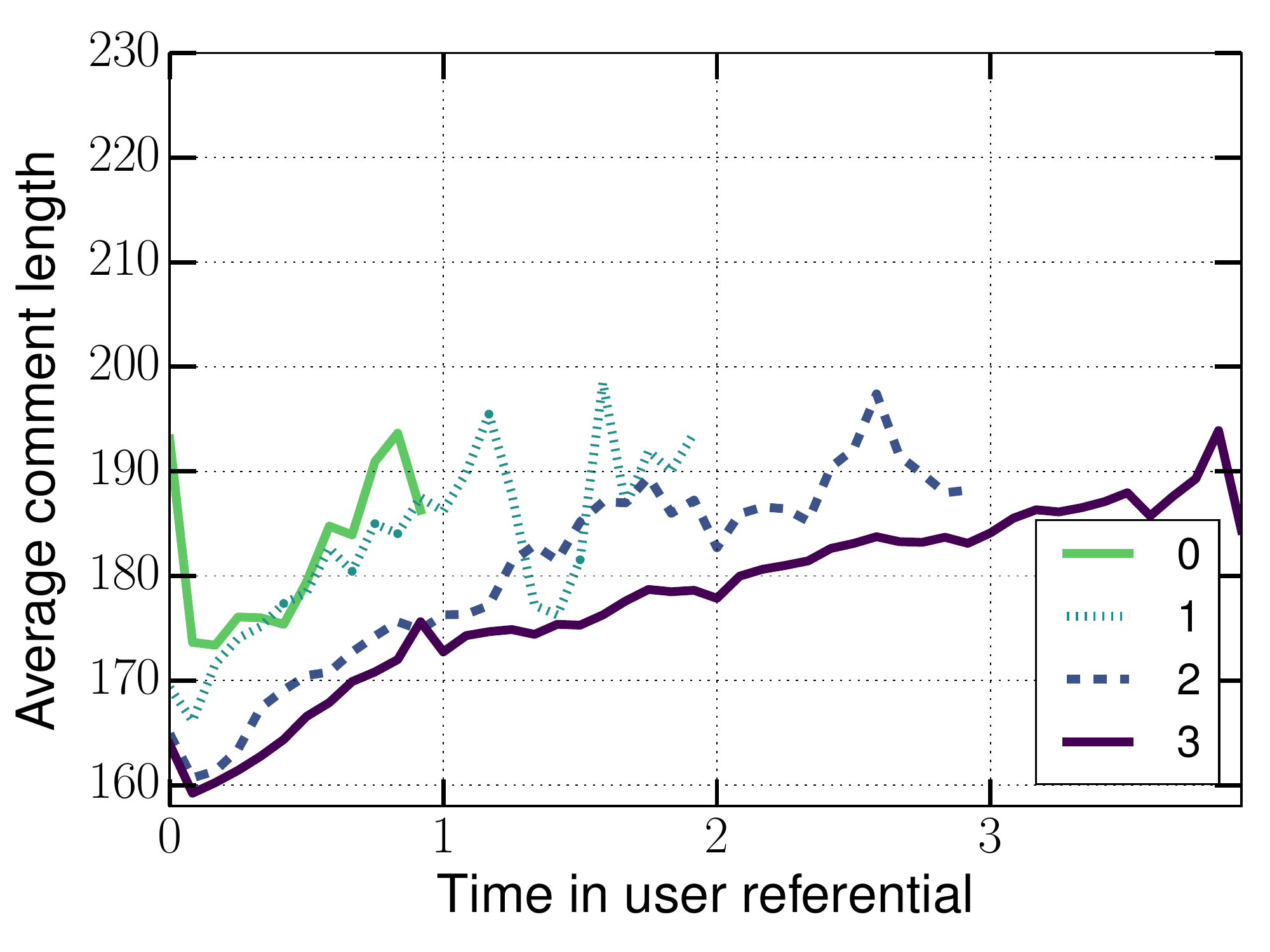}{2011 cohort}
\subimage[width=0.31, scale=0.29]{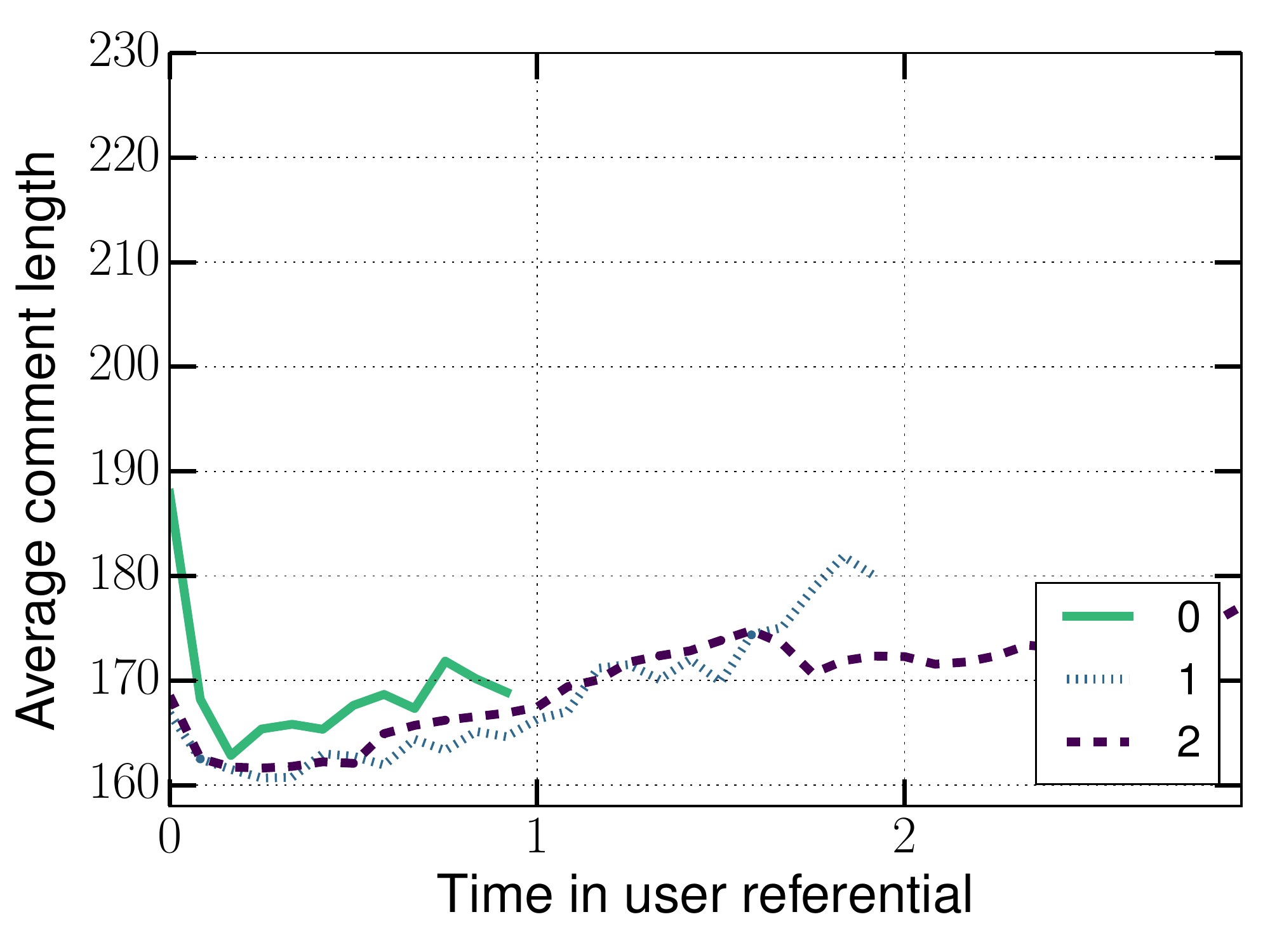}{2012 cohort}
\caption{Figure (a) shows the average comment length over clock time and Figure (b) from the user-referential time. Both figures show the cohorted trends.  The overall average length per comment decreases over time, although for any individual cohort, it increases after a sharp initial drop. Figures (c), (d) and (e), similar to Figure~\ref{fig:avr_posts_per_user_for_surviving_year}, show the monthly average comment length for active users in the cohorts of 2010, 2011 and 2012, segmented by the number of years that the user survived in the network.  Opposite the analysis for average posts, which showed that low-activity users were the first to leave Reddit, here, people who start out as longer commenters are \textit{more} likely to leave.}
\label{fig:comment_length}
\end{figure*}

Activity as measured by the average number of posts per user is one proxy for user effort.  Comment length can also be considered as a proxy for user effort in the network.  Users that type more put more of their time in the network, contribute with more content, and might create stronger ties with the community. Thus, we put forward the following question (\textbf{RQ2}): how does comment length change in the community over time, both overall and by cohort?

\subsection{Comment length drops over time}

Figure~\ref{fig:comment_length}a shows the overall comment length in Reddit over time (the darker line) and the overall length per cohort. 
Based on the downwards tendency of the overall comment length in Figure~\ref{fig:comment_length}a, one might hypothesize that users' commitment to the network is decreasing over time (\textbf{H3}), or that there is some community-wide norm toward shorter commenting (\textbf{H4}). 

However, this might not be the best way to interpret this information. Figure~\ref{fig:comment_length}b shows the comment length per cohort in the user referential time. An important observation here is that younger users start from a lower baseline comment length than older ones. Considering the fact that Reddit has experienced exponential growth, the overall average for Figures \ref{fig:comment_length}a and \ref{fig:comment_length}b is heavily influenced by the ever-growing younger generations, who are more numerous than older survivors and who post shorter comments. 

\subsection{Simpson's Paradox: the length also rises}

Let us go back to Figure~\ref{fig:comment_length}a, which shows the overall average comment length on Reddit over time. We see a clear trend towards declining length of comments in the overall line (the black line that averages across all users). This could be a warning sign for Reddit community managers, assuming longer comments are associated with more involved users and healthier discussions. A data analyst looking at these numbers might think about ways to promote longer comments on Reddit. 

However, Figure~\ref{fig:comment_length}b shows that average comment length increases over time for every cohort. While later cohorts start at smaller comment length, after an initial drop, on average all cohorts write longer comments over time.  This is puzzling: when each of the cohorts exhibits a steady increase in their average comment length, how can the overall mean comment length decrease?  This anomaly is an instance of the Simpson's paradox \cite{simpson1951}, and occurs because we fail to properly condition on different cohorts when computing mean comment length. 

\begin{table}[!tb]
\centering
\tabcolsep=0.07cm
\singlespacing
\fontsize{9pt}{10.5pt}\selectfont
\begin{tabular}{|c|c|c|c|c|c|c|c|c|c|}
\cline{2-9}
\multicolumn{1}{c|}{} & \multicolumn{8}{c|}{Cohorts} \\ \hline
Year & 2007 & 2008 & 2009 & 2010 & 2011 & 2012 & 2013 & 2014 & Overall\\ \hline
2007 & 220 & - & - & - & - & - & - & - & 220 \\ \hline
2008 & 208 & 198 & - & - & - & - & - & - & 204 \\ \hline
2009 & 224 & 204 & 201 & - & - & - & - & - & 208 \\ \hline
2010 & 223 & 204 & 189 & 184 & - & - & - & - & 193 \\ \hline
2011 & 233 & 211 & 199 & 184 & 167 & - & - & - & 182 \\ \hline
2012 & 241 & 221 & 212 & 197 & 173 & 167 & - & - & 178 \\ \hline
2013 & 244 & 225 & 214 & 199 & 177 & 167 & 164 & - & 174 \\ \hline
2014 & 246 & 229 & 217 & 204 & 183 & 172 & 165 & 176 & 176 \\ \hline
\end{tabular}
\caption{Evolution of the average throughout the years for each cohort. Each column here is one cohort and each line is one year in time. Cohorts start generating data in their cohort year, therefore the upper diagonal is blank. On the right column we see the overall average for all users.}
\label{tab:simpson}
\end{table}

Table~\ref{tab:simpson} provides some clues to what might be going on. When we move down the rows, we observe an increasing tendency in each cohort column. It means that the average comment length increases for these users. However, when we move right through the columns, people in later cohorts tend to write less per comment. If we were to average each row, we would still get an overall increasing comment length per year, but that is not what we see in the overall column. What happens here is that the latter cohorts have many more users than earlier ones. Since their numbers increase year by year, we have a much larger contribution from them towards comments, compared to users of earlier cohorts. This uneven contribution leads to the paradox we observed in Figure~\ref{fig:comment_length}a. 

Without the decision to condition on cohorts, one would have gathered an entirely wrong conclusion. People are not writing less as they survive, contra (\textbf{H3}).  Rather, those who tend to write less are joining the community in much larger numbers.  Why later users write less is an open question we speculate about later in the discussion and future work section.

\begin{figure*}[!tb]
\centering
\subimage[width=0.48, scale=0.42]{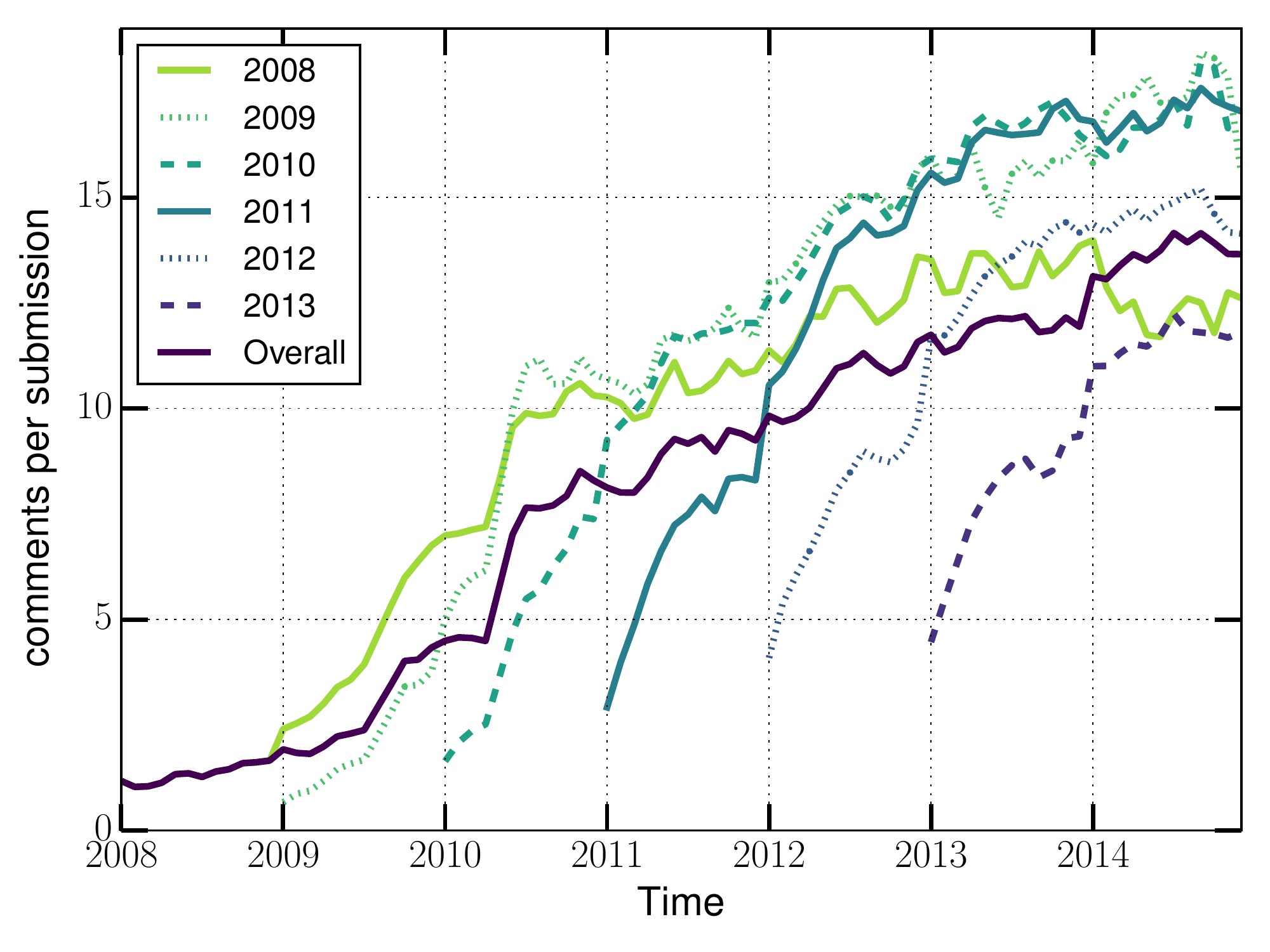}
\subimage[width=0.48, scale=0.42]{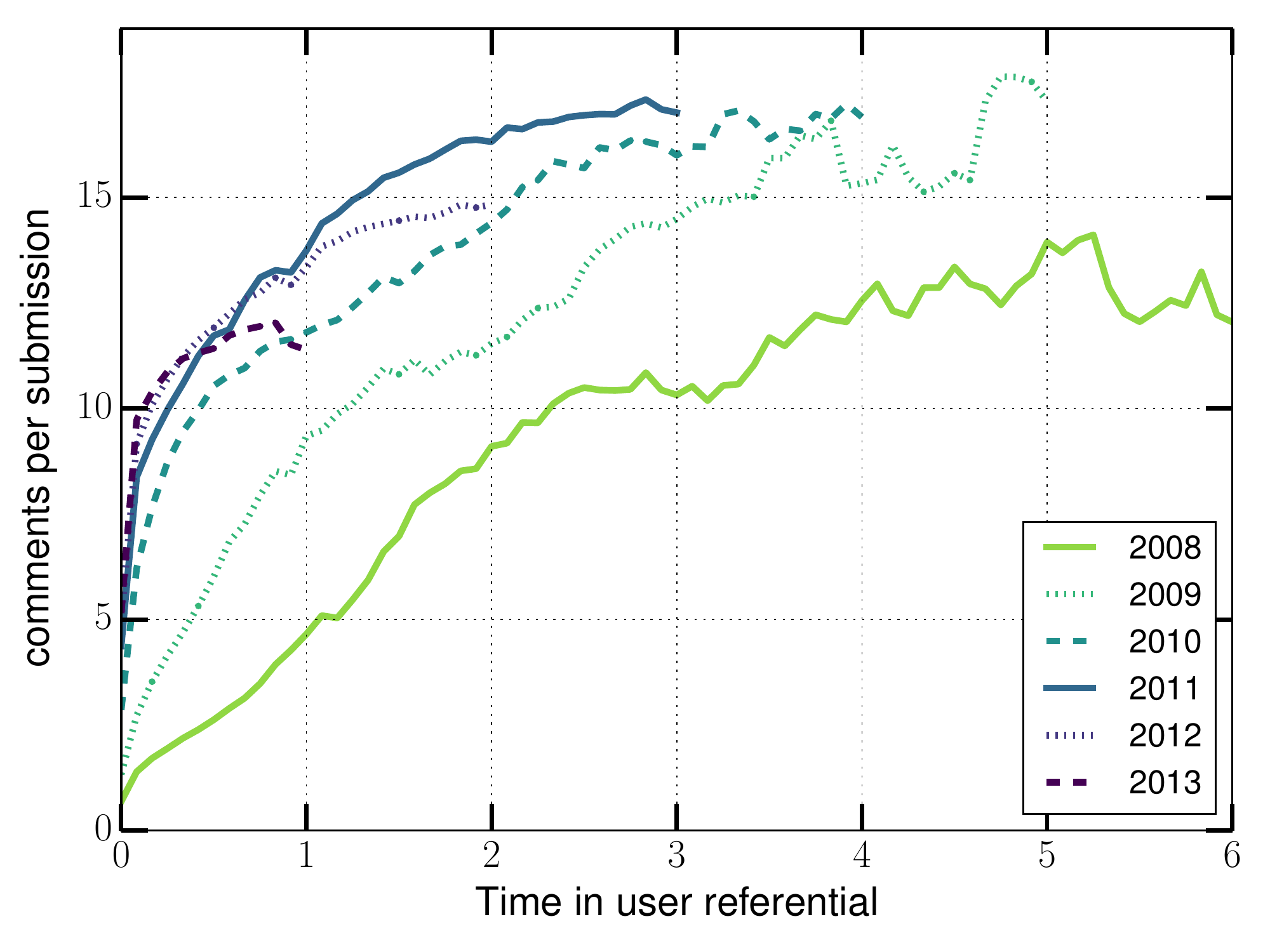}
\subimage[width=0.23, scale=0.23]{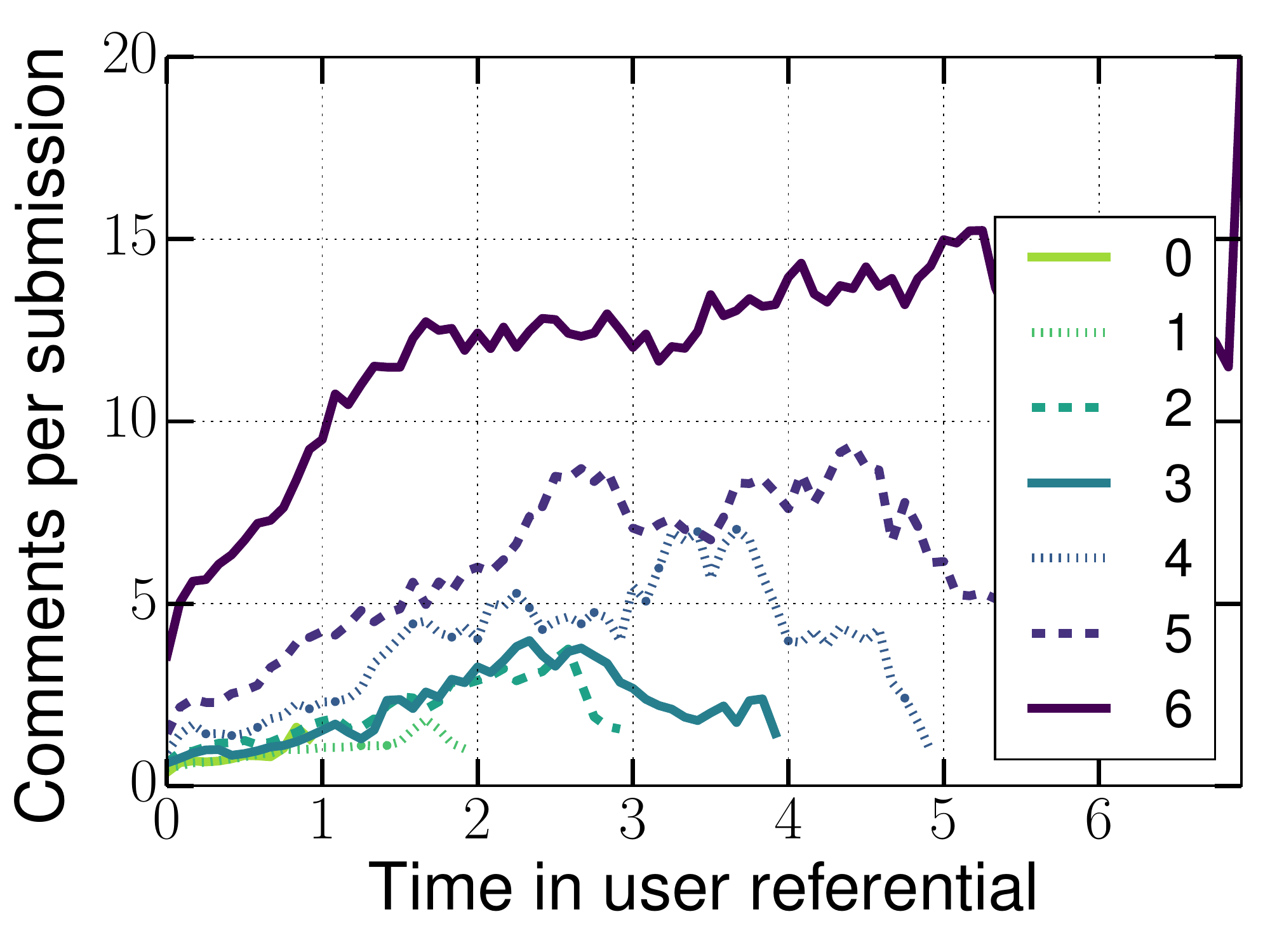}{2008 cohort}
\subimage[width=0.23, scale=0.23]{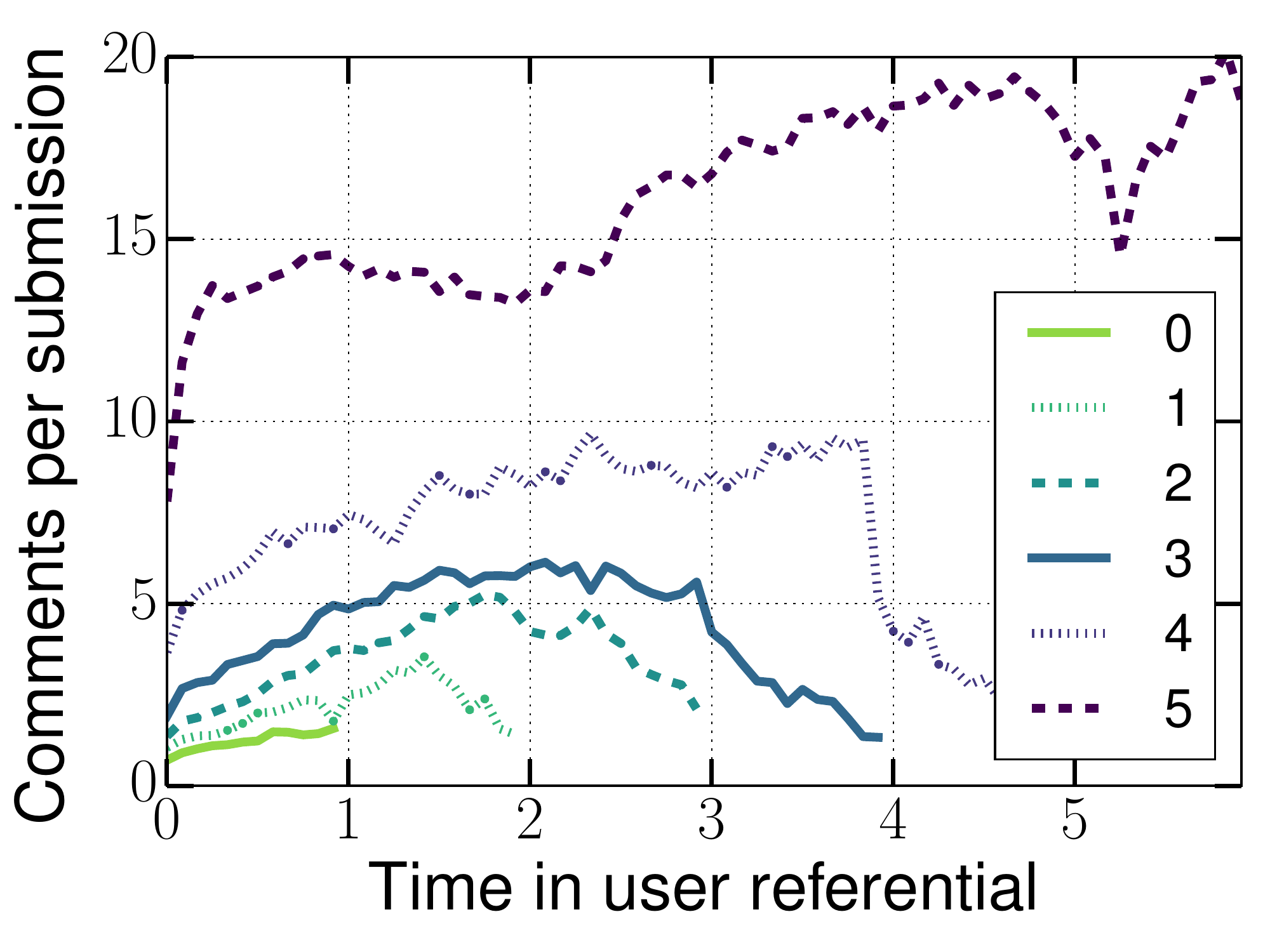}{2009 cohort}
\subimage[width=0.23, scale=0.23]{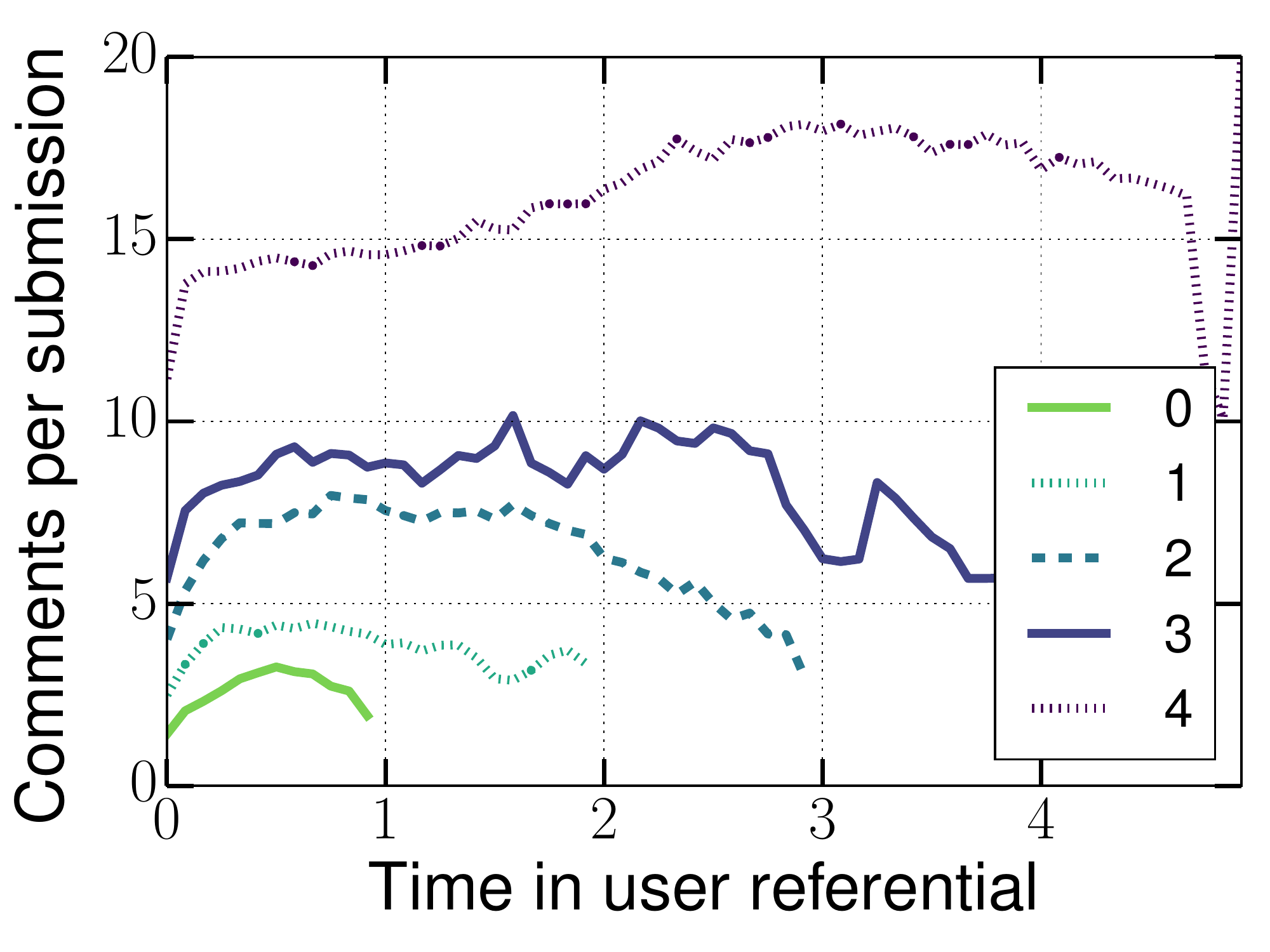}{2010 cohort}
\subimage[width=0.23, scale=0.23]{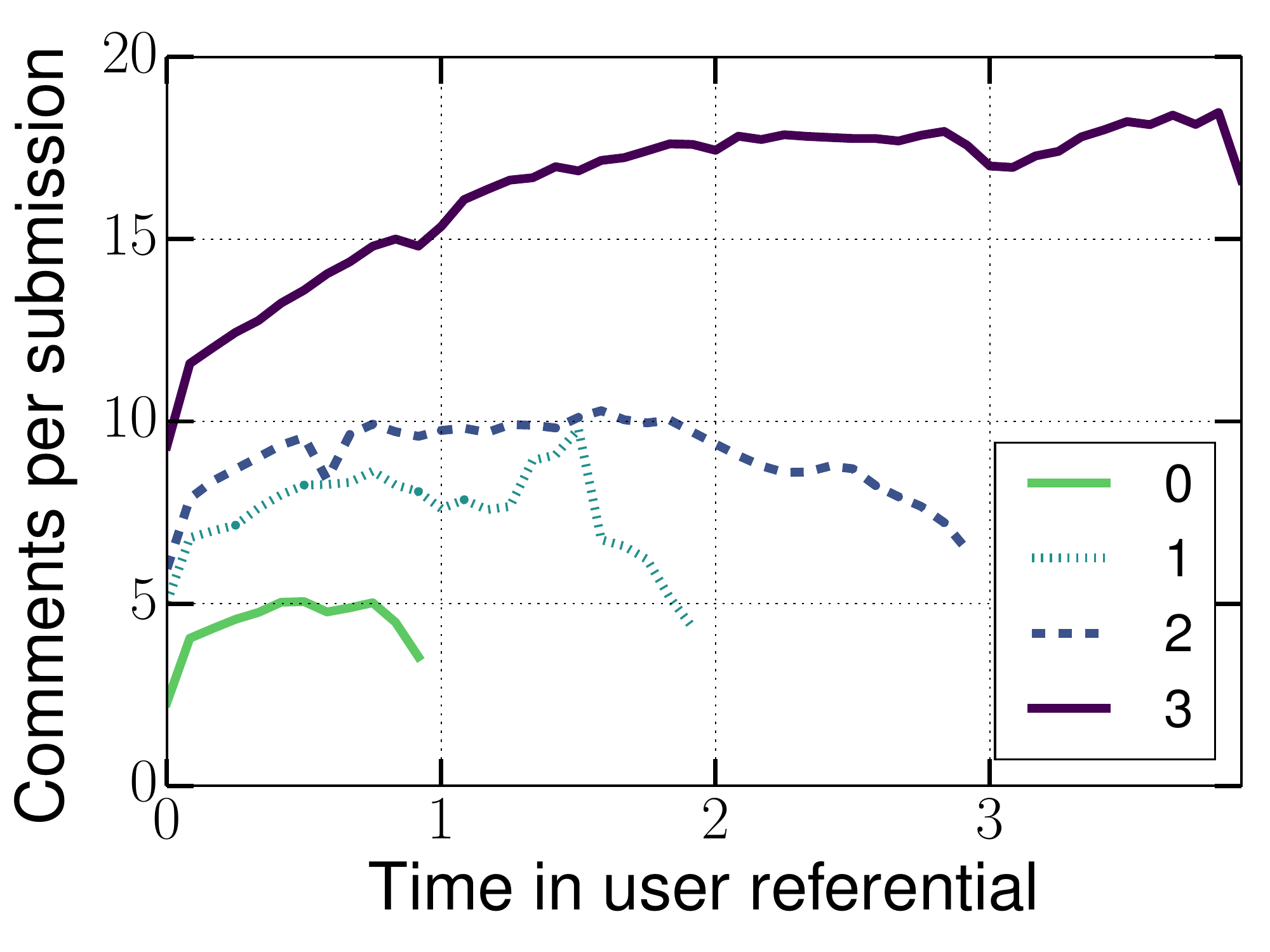}{2011 cohort}
\caption{Figure (a) shows the average comment per submission ratio over clock time for the cohorts and the overall average. Figure (b) shows the average comment per submission from the user-referential time for the cohorts. Figures (c), (d), (e) and (f), similarly to Figure~\ref{fig:avr_posts_per_user_for_surviving_year}, shows the 2008, 2009, 2010, and 2011 cohorts, segmented by the number of years a user in the cohort survived.  As with average posts per month, users who stay active longer appear to start their careers with a relatively higher comments per submission ratio than users who abandon Reddit sooner.  Unlike that analysis, however, the early 2008 cohort ends up below the later cohorts in Figure (b).}
\label{fig:comments_submissions}
\end{figure*}

\subsection{New users burn brighter}
As with the number of posts per user, we cannot say if the increase in the curves seen in \ref{fig:comment_length}b is due to lower-effort users dying first or because users are writing more as they live longer.  The sub-cohort analysis in \ref{fig:comment_length}c allows us to make two observations toward this question.  First, \textit{comment length does increase inside of each cohort}, no matter how long the user survives.  Second, as a general trend, \textit{users that make longer comments inside of each cohort die faster}. This is quite surprising, given that we would expect people to put less effort when they are more likely to stop using the network.

\section{Kinds of contributions}

In addition to questions of effort, the online community literature also often asks what sorts of activities users engage in, for instance, to categorize users into roles they play in the community \cite{Welser2011}. As with comment length, we propose the following research quetion (\textbf{RQ3}): how do users' activities change in the community over time, both overall and by cohort?

\subsection{Over time, responsiveness increases}
Consider the case of Usenet: people who never start threads and only respond play the role of answerer, while there are other roles that include fostering discussion \cite{Welser2007}.  These might naturally map onto people who primarily comment and who primarily submit in Reddit, respectively.  Submissions can be considered new content that an author generates, while comments can be considered as contributions toward existing content from another author.

Since the total number of comments always surpasses the number of submissions, we compute a user's ratio of comments per submission as a rough measure of the kinds of contributions they make.  Figure~\ref{fig:comments_submissions}a shows the overall and cohorted evolution of comments per submission from 2008 to 2013.  Users who most prefer commenting to submitting come from 2009 to 2011, while over time the average ratio of comments to submissions increases both overall and per-cohort for active users.

Again, we analyze our data from the user-time referential, as seen in Figure~\ref{fig:comments_submissions}b. It shows a clear pattern for users in earlier cohorts to have a lower comment per submission ratio than users in later cohorts, given that they both survived the same amount of time.  Surviving users from later cohorts also exhibit a more rapid increase in comments per submission than those from earlier cohorts.  In particular, the 2008 and 2009 cohorts increase much more slowly over time than those from 2010 onwards; later cohorts are more similar (although the 2012 and 2013 cohorts may level off lower than 2011 based on the limited data we have). 

\subsection{Comment early, comment often}

Figures~\ref{fig:comments_submissions}c-f shows the cohorts from 2008 to 2011 segmented by surviving year.  Three interesting observations arise from these data.  First, we see that just as in the analysis of average posts per user, the users who survive the longest in each cohort are the ones who hit the ground running.  They start out with a high comment-to-submission ratio relative to users in their cohort who abandon Reddit more quickly.  This suggests that both the count of posts and the propensity to comment might be a useful early predictor of user survival.

Second, and unlike the case for average post length, surviving users' behavior changes over time.  For post length, Figure~\ref{fig:avr_posts_per_user_for_surviving_year} shows that even the most active users come in at a certain activity level and stay there, perhaps even slowly declining over time.  Here, Figures~\ref{fig:comments_submissions}c-f show that the ratio of comments to submissions increases over time.  Combined with the observation that overall activity stays steady, this suggests that the ratio is changing because people \textit{substitute} making their own submissions for commenting on others' posts.

Finally, this increase is most pronounced in the earlier cohorts of 2008 and 2009, with ratios more than doubling over their first year, much more than for later cohorts.

\section{Discussion and Future Work}

In this section we discuss some of the processes that might explain our observations, and how they connect to other literature.  We're not arguing here that we know the answers; instead, we see these as interesting avenues for future work.  

\subsection{Why are newer ``active'' users less so?}

We have seen that users from later cohorts have a lower posting average than in earlier cohorts. 
One plausible explanation is that users self-select: users that find Reddit early in its life are also more likely than average to be those who will be attracted to it. Previous work has shown that online book reviews have a self-selection bias, where people who are more likely to like (or promote) the book review it earlier, leading to a positive early bias in an item's life \cite{Li2008}.  In Reddit's case, this would mean that the mixture of users joining in the early stage of the community would be disproportionately likely to be the most active ones and the latter ones are more likely to be less active; several of our results support this explanation.

Another plausible hypothesis for later cohorts having a higher number of less active users could be that, over time, Reddit has accumulated an increasing number of valuable-but-small/niche communities.  The increased diversity might support a wider set of users in getting value, explaining the increased survival percentage.  The niche/smaller nature of newer communities might provide fewer opportunities to both submit and comment, explaining the lower average activity for surviving users. 

A third hypothesis is that Reddit overall is becoming more about consumption and voting on content rather than producing it.  Older users with contribution norms continue to contribute; newer users tend to provide audiences and feedback.  High-resolution voting data could be a real boon in understanding if this is true.

\subsection{Why are comments getting shorter?}

We also observed that overall, comment lengths are getting shorter over time.  
One hypothesis is that users' behavior is being shaped by an ``initial value problem''---that as users join the network, they tend to produce content according to the norms of what they see \cite{Kooti2010, Danescu-niculescu-mizil2013}.  
Figure~\ref{fig:comment_length}a presents some support for this hypothesis: the initial month of each cohort year, which consists of data only from users who joined in that month, is quite close to the overall line from the prior month.  

Another hypothesis advanced by community members\cite{RedditHypo1} is that Reddit's karma system favors shorter comments.  That is, people can get more upvotes for a given amount of effort by writing more, shorter comments.  This could be directly measured even with the available data, and might be the start of a very interesting line of future work around modeling strategic posting and attention distribution behavior in Reddit. 

\subsection{Why do comments per submission increase?}

We also saw that comments per submission increase over time for surviving users, especially for users who join earlier.

One process hypothesis is that this is because early in Reddit's life, there simply weren't as many submissions to comment on, meaning that people who wanted to be active contributors more or less had to submit in order to do so. 
As the community grew, more content became available to comment on; those comments in turn provide additional opportunities for commenting.  In this reading, the value and ease of commenting has increased over time, making it a more common behavior. 

This question of ease and value might be more general, and tie to our observations about self-selection and karma accumulation.  Most users in social networks are known to be lurkers: seeking information and observing, rather than contributing content \cite{Rafaeli2004, Nonnecke2000}. Consumption in Reddit is valuable and easy, and some contributions are easier than others: reading is easier than voting; voting is easier than commenting; commenting is easier than submitting.  Only users for whom finding and submitting comments is relatively easy or relatively valuable are likely to be frequent submitters or ``power users'' \cite{Panciera2009, Kittur2007}. We suspect such users are more likely to be ones who found Reddit earlier, when it was relatively small, and stuck with it.

\subsection{Limitations and Future Work}

In this paper we focused our attention on visible behavior attributable to specific users, which in this dataset meant submissions and comments.  As with many analyses that focus on visible behavior, this means we miss important phenomena.  In particular, we discount lurkers despite their known importance as audience members \cite{Nonnecke2003} and potential future contributors \cite{Ridings2006}.  Many lurkers likely vote, and thus lurking may be even more important in a context like Reddit where votes affect content visibility and provide explicit markers of attention and reputation.  

However, the dataset does not have information on individual voters or timestamps, just the aggregate number of votes a post had received at the time of the crawl, making it impossible to use them as activity measures for specific users.  The existing voting data might be much more useful, however, in addressing questions that involve predicting a given user's future behavior based on how other users respond to a user's early contributions \cite{Joyce2006,Sarkar2012}.

Focusing on visible activity can lead to blind spots in other places, as well.  In particular, our emphasis on active users led us to ignore questions of survival, leaving, and rejoining.  This was a reasonable view of the community based on the questions we were asking, but our results should all be interpreted in the context of ``given the set of active users at any given time''.  Applying these results to questions that require considering all users would be a mistake.  

We did, implicitly, consider survival in the analyses that broke cohort down by survival time; more generally, we see careful thinking about what it means to ``survive'' in a community as an interesting problem in its own right.  Many analyses assume that a gap of some time period implies that a user has left, or that users ``die'' on their last visible day of activity.  However, long gaps are common in real behavior.  People temporarily quit social media all the time \cite{Baumer2013}, and in Wikipedia, the practice of leaving temporarily is so common it has a name: ``wikibreak''.    Rather than an annoying right censorship statistical problem, this question of what it means when contributors to a community start and stop might pose a much more central issue, as a community's survival might not depend only in its ability to attract and retain users, but also in the ability to ``resurrect'' old users and leverage ``bursty'' ones.

\looseness=-1

Further, One of our assumptions was that one account is associated with one user. This might not be the case, as more than one user can share the same account \cite{Lampinen2014} or one user can have multiple accounts \cite{Bergstrom2011}.  Multiple accounts can have many functions, including making points someone doesn't want connected with their main identity, trolling or harming other users or the community, or simulating users who agree with a main identity (``sock puppets'').  While we think this is not the main driver of our results, this should be checked in future work---and sockpuppet detection and account deanonymization is an interesting question in its own right.

Finally, focusing on visible activity can also lead to blind spots around deleted content or communities.  At least in Reddit, activity from users is marked with a username of ``[deleted]'', which we discovered after realizing that one author had millions of comments(!), and that allowed us to consciously choose to exclude that data.  However, in some contexts, such as Wikipedia articles that are deleted, that activity is invisible as edit behavior on those articles does not show up in many data dumps.  Such invisible activity might be important in understanding either individual users or the community. 

\section{Conclusions}

This work highlights the importance of taking time into consideration when analyzing users' evolution in social networks. We do so by cohorting the users based on their creation year. Although simple, this approach provides a number of insights that would be missed by straightforward aggregate analysis methods.  We also analyze the evolution of users and communities from a shifted time referential: considering the time of an action in relation to the user creation date. This also reveals unexpected phenomena that we would otherwise not notice.

While analyzing how the amount of posting changes over time (\textbf{RQ1}), we found that user posting activity for surviving Reddit users is actually significantly higher than a naive average would suggest, that older users who survive are considerably more active than younger survivors, and that these newer users are unlikely to catch up (\textbf{RQ1a}).  Controlling for survival provided evidence for hypothesis (\textbf{H2}), that users have a stable level of posting activity over time (with slightly decreasing patterns).  Further, the percentage of surviving but low-activity users is increasing in the younger cohorts 

When looking at changes in comment length over time \textbf{RQ2} as a proxy for users' effort, we found that while the overall average in Reddit seems to decrease, users actually write longer comments as they survive, no matter when they join.  However, later cohorts of users that joined the network are writing smaller comments; their greater number leads to an instance of Simpson's paradox, where the overall average decreases while the series for each individual cohort increases. 

Finally, we analyze whether users change their commenting versus submission behavior over time (\textbf{RQ3}). 
We found that users with a higher initial comment to submission ratio survive longer on average, and that this ratio increases for surviving users, particularly for earlier cohorts.  This isn't because activity rises overall, as posting activity remains stable; instead, it suggests that longer-term users substitute commenting for submissions. 

An important remark of this paper is how different demographics of users joining and leaving a network play a significant role in shaping the average user behavior. Failing to account for these might limit our interpretation of the data (\textbf{H1}, \textbf{H3} or \textbf{H4}) and lead to wrong conclusions.

Both our and work and its limitations suggest fruitful directions for better understanding of users' evolution in both Reddit and online communities in general, directions we hope inspire other work in this area.  

\section{Acknowledgments}
The authors are grateful to FAPESP grant \#2011/50761-2, CAPES grant \#99999.009323/2014-07, NAP eScience - PRP - USP, Google, NSF IIS-1422484, and RepNerv.

\bibliographystyle{abbrv}
\bibliography{www2016}

\end{document}